\documentclass[thmsa,11pt]{article}
\usepackage{amsmath}
\usepackage{amsfonts}
\begin{document}

\title{\hfill {\small IFUM-913-FT \medskip }\\
\textbf{Thermodynamics of Einstein-Born-Infeld black holes with negative
cosmological constant}}
\author{Olivera Mi\v{s}kovi\'{c}$\,^{a}$ and Rodrigo Olea$\,^{b}$\medskip \\
{\small \emph{$^{a}$Instituto de F\'{\i}sica, P. Universidad Cat\'{o}lica de
Valpara\'{\i}so, Casilla 4059, Valpara\'{\i}so, Chile.}} \\
{\small \emph{$^{b}$INFN, Sezione di Milano, Via Celoria 16, I-20133,
Milano, Italy.}}\\
{\small \texttt{olivera.miskovic@ucv.cl, rodrigo.olea@mi.infn.it,}}}
\date{}
\maketitle

\begin{abstract}

We study the thermodynamics associated to topological black hole
solutions of AdS gravity coupled to nonlinear electrodynamics
(Born-Infeld) in any dimension, using a background-independent
regularization prescription for the Euclidean action given by
boundary terms which explicitly depend on the extrinsic curvature
(Kounterterms series). A finite action principle leads to the
correct definition of thermodynamic variables as Noether charges,
which satisfy a Smarr-like relation. In particular, for the
odd-dimensional case, a consistent thermodynamic description is
achieved if the internal energy of the system includes the vacuum
energy for AdS spacetime.

\end{abstract}

\section{Introduction}

A remarkable feature of nonlinear Born-Infeld electrodynamics \cite%
{Born-Infeld} is that it is able to describe a classical theory of charged
particles with finite self-energy and an electric field that is regular at
the origin. Due to a screening effect on the electric field for small
distances, its source can be interpreted as an extended object with an
effective radius $r_{\text{eff}}=q^{2}/b^{2}$, where $q$ is related to the
electric charge of the particle and $b$ is the Born-Infeld coupling.

The regulated behavior of this electromagnetic theory makes a
Born-Infeld-like action a sensible candidate for a gravitational theory
\cite{Deser-Gibbons, Klemm}. Indeed, such action arises naturally in string
theory as it governs the dynamics of D-branes
\cite{Fradkin-Tseytlin,Tseytlin,Leigh,Callan-Maldacena,Gibbons}.

When Born-Infeld electrodynamics is coupled to anti-de Sitter (AdS)
gravity, charged black holes exhibit thermodynamic properties
similar to the ones of Reissner-Nordstr\"{o}m AdS solutions
\cite{Chamblin-Emparan-Johnson-Myers,Peca-Lemos,Lemos-Zanchin}
showing a phase structure isomorphic to van der Waals-Maxwell
liquid-gas system (for the four-dimensional Einstein-Born-Infeld
(EBI) AdS case, see \cite{Fernando,Fernando-Krug}). Asymptotically
(A)dS black hole configurations in this theory were found in an
arbitrary dimension and their thermodynamics was discussed in
Refs.\cite{Dey,Cai-Pang-Wang}. Electrically charged rotating black
branes in EBI AdS gravity were studied in \cite{Dehghani-Sedehi}.
Solutions of nonlinear BI like (Hoffman-Infeld) electromagnetism
coupled to Lovelock gravity were found in
\cite{Aiello-Ferraro-Giribet}. Finally, the attractor mechanism for
spherically symmetric extremal black holes in EBI-dilaton theory of
gravity in four dimensions with cosmological constant was
investigated in \cite{XGao}.

As in Einstein-Maxwell gravity, the black hole entropy $S$ is a quarter of
the horizon's area (in fundamental units), and the variations of the
solution parameters obey the First Law of thermodynamics%
\begin{equation}
\beta dU=dS+\beta \Phi d\mathcal{Q\,},  \label{First Law}
\end{equation}%
where $U$ is the internal energy, $\beta $ is the inverse of the black hole
temperature, $\Phi $ is the gauge potential measured at infinity respect to
the horizon and $\mathcal{Q}$ is the \emph{thermodynamic }electric charge.

In general, it is possible to show that the first law holds simply assuming
that the thermodynamic variables whose variations appear in Eq.(\ref{First
Law}) are the conserved charges of the theory. However, the thermal
properties of black holes can only be understood when one identifies the
Euclidean path integral with the thermal partition function. In the
semiclassical approximation, the partition function is given by the
exponential of the classical Euclidean action $I_{clas}^{E}$,
\begin{equation}
Z=e^{-I_{clas}^{E}},  \label{Z}
\end{equation}
and therefore, for spacetimes with AdS asymptotics, $I_{clas}^{E}$ needs to
be regulated to cancel the divergences that appear in the asymptotic region.
In doing so, the thermodynamic information is encoded in the finite part the
Euclidean action $I^{E}$, which obeys the Smarr relation
\begin{equation}
I^{E}=\beta U-\beta \Phi \mathcal{Q}-S\,.  \label{Smarr}
\end{equation}
Background-substraction methods for asymptotically AdS spacetimes are useful
to extract a finite value from $I^{E}$, considering the Euclidean action for
the black hole minus the same functional evaluated for the corresponding
vacuum solution. The result satisfies Eq.(\ref{Smarr}) and it is equivalent
to integrate out the first law for global AdS ($U=\mathcal{Q}=0$) as the
background. The procedure, however, does not guarantee the existence of such
background for a complex enough gravitational configuration because it is
not always possible to embed an arbitrary solution into a given reference
spacetime.

The identification of the thermodynamic variables as asymptotic charges may
be more subtle in the context of the AdS/CFT correspondence \cite{AdS/CFT}.
In this framework, the standard regularization method consists on the
addition of counterterms which are covariant functionals of the boundary
metric and intrinsic curvature, constructed by a systematic procedure known
as holographic renormalization
\cite{Henningson-Skenderis,deHaro-Skenderis-Solodukhin}. The counterterms cancel
the divergences in both conserved charges and Euclidean action in a
background-independent fashion
\cite{Balasubramanian-Kraus,Emparan-Johnson-Myers}. The novel feature is the
appearance of a nonvanishing zero-point energy $E_{vac}$ for global AdS
spacetime in the odd-dimensional case, which matches the Casimir energy of a
boundary CFT, and it is helpful to realize explicit examples of the
gravity/gauge duality.

It is clear that the first law (\ref{First Law}) is insensitive to the
existence of a vacuum energy for asymptotically AdS spacetimes. What is far
from evident is whether the Smarr relation is still valid for an internal
energy corresponding to the total energy of the system in odd-dimensional
gravity with negative cosmological constant. Employing Dirichlet
counterterms, one is able to check, on a rather case-by-case basis, the
consistency of the formula (\ref{Smarr}), only if the internal energy is
shifted as $U=M+E_{vac}$ with respect to the Hamiltonian mass $M$. That
implies that the regulated Euclidean action also appears shifted
accordingly, respect to the value obtained in a background-substraction
procedure.

An explicit proof of the Smarr formula in an arbitrary dimension faces the
problem of finding a general counterterm action, which is still unknown for
a high enough dimension \cite{Papadimitriou-Skenderis}. It also presupposes
that one can isolate the contribution from the stress tensor to $E_{vac}$ so
as to write down a covariant formula only for that part. However, the fact
that such formula has not been found prevents from casting the shift in
$I^{E}$ as $\beta E_{vac}$ for an arbitrary asymptotically AdS solution.

An alternative regularization scheme for gravity with AdS
asymptotics, known as Kounterterms method, has been recently
proposed \cite{Olea-KerrBH,Olea-K}. This prescription considers
supplementing the action by boundary terms that are a given
polynomial of the intrinsic and extrinsic curvatures. A remarkable
feature of this approach is its universality as --for a given
dimension-- the Kounterterms series preserve its form for
Einstein-Gauss-Bonnet AdS \cite{Kofinas-Olea-EGB} and even for any
Lovelock gravity with AdS asymptotics \cite{Kofinas-Olea}.
Furthermore, because of a profound relation to topological
invariants and Chern-Simons forms, the explicit form of the boundary
terms can be written down in any dimension. As a consequence,
background-independent formulas for the conserved charges which, on
the contrary to what happens in the standard counterterm procedure,
can be shown in all dimensions. Background-independence is
particularly relevant when global AdS spacetime possesses a
non-vanishing energy. It has been shown in Ref.\cite{Olea-K} that a
covariant formula for the vacuum energy in an arbitrary odd
dimension can indeed be obtained using Kounterterms method.

In this paper, we study black hole thermodynamics in EBI AdS gravity
as an application of the regularization prescription described
above. First, we are able to identify the thermodynamic variables of
the system as Noether charges, which are rendered finite employing
Kounterterm series. Second, the direct evaluation of the Euclidean
action allows us to recognize the contribution from radial infinity
as the corresponding conserved quantities and to show that they
verify a Smarr-type formula. Finally, in the odd-dimensional case,
we show that this picture is consistent only if the internal energy
$U$ includes the vacuum (Casimir) energy for AdS spacetime.


\section{Action and equations of motion}

We consider the Einstein-Born-Infeld gravity with negative cosmological
constant in $D=d+1$ dimensions,
\begin{equation}
I_{reg}=\int\limits_{M}d^{D}x\,\sqrt{-g}\,\mathcal{L}+c_{d}
\int\limits_{\partial M}d^{d}x\,B_{d}\,,  \label{I}
\end{equation}
where the bulk Lagrangian density has the form
\begin{equation}
\mathcal{L}=-\frac{1}{16\pi G}\,\left[ \hat{R}-2\Lambda +4b^{2}\left( 1-
\sqrt{1+\frac{F^{2}}{2b^{2}}}\right) \right] \,,  \label{EH-BI}
\end{equation}
and the boundary term $B_{d}$ shall be discussed below. Hatted curvatures
refer to the full spacetime, which is endowed with the metric tensor $g_{\mu
\nu }$.\ The first two terms in (\ref{EH-BI}) correspond to the
Einstein-Hilbert action with negative cosmological constant (for conventions
see Appendix \ref{Conventions}),
\begin{equation}
\Lambda =-\frac{\left( D-1\right) \left( D-2\right) }{2\ell ^{2}}\,,
\end{equation}
and $G$ is the gravitational constant. The second part is the Born-Infeld
term, where the parameter $b$ (with dimension of mass) is related to the
string tension $\alpha ^{\prime }$ as $b=\frac{1}{2\pi \alpha ^{\prime }}$
and $F_{\mu \nu }=\partial _{\mu }A_{\nu }-\partial _{\nu }A_{\mu }$ is the
field strength associated to the Abelian gauge field $A_{\mu }$. We also
denote $F^{2}=g^{\mu \lambda }g^{\nu \rho }F_{\mu \nu }F_{\lambda \rho }$.
In the limit $b\rightarrow \infty $, the Born-Infeld term recovers Maxwell
electrodynamics, whereas in the limit $b\rightarrow 0$, it vanishes.

In $D=4$, apart from $F^{2}$, one can construct another quadratic
invariant, $\tilde{F}^{\mu \nu }F_{\mu \nu }$, which involves the
dual of the field strength $\tilde{F}$. This term, when added to the
action, is locally a boundary term and thus, it does not change the
bulk dynamics. However, nonlinear electrodynamics action may include
this quadratic invariant in a nontrivial form, on the condition that
the Maxwell limit is recovered \cite{Wiltshire,Rasheed,Breton}.
Unfortunately, because of the dependence on $\tilde{F}$, such action
cannot be extended to higher dimensions than four.

Varying the action with respect to the metric $g_{\mu \nu }$ and the gauge
field $A_{\mu }$ produces the equations of motion plus a surface term,
\begin{equation}
\delta I_{reg}=\frac{1}{16\pi
G}\int\limits_{M}d^{D}x\,\sqrt{-g}\left( \left( g^{-1}\delta
g\right) _{\mu }^{\nu }\,\mathcal{E}_{\nu }^{\mu }+\delta A_{\mu
}\,\mathcal{E}^{\mu }\right) +\int\limits_{\partial
M}d^{d}x\,\Theta \text{\thinspace },  \label{var_Ibulk}
\end{equation}
where
\begin{eqnarray}
\mathcal{E}_{\nu }^{\mu } &\equiv &\hat{R}_{\nu }^{\mu }-\frac{1}{2}
\,\delta _{\nu }^{\mu }\,\hat{R}+\Lambda \,\delta _{\nu }^{\mu }-T_{\nu
}^{\mu }\,,  \label{eom1} \\
\mathcal{E}^{\mu } &\equiv &-\hat{\nabla}_{\nu }\left(
\frac{4F^{\mu \nu }}{\sqrt{1+\frac{F^{2}}{2b^{2}}}}\right) \,.  \label{eom2}
\end{eqnarray}
Here $\hat{\nabla}_{\mu }$\ denotes covariant derivative with respect to the
Christoffel connection $\hat{\Gamma}_{\mu \nu }^{\alpha }$,\ whereas the
Born-Infeld electromagnetic stress tensor is given by
\begin{equation}
T_{\nu }^{\mu }=2b^{2}\delta _{\nu }^{\mu }\left( 1-\sqrt{1+
\frac{F^{2}}{2b^{2}}}\right) +\frac{2F^{\mu \lambda }F_{\nu \lambda }}{\sqrt{1
+\frac{F^{2}}{2b^{2}}}}\,.
\end{equation}
For a large value of the parameter $b$, we have
\begin{equation}
T_{\nu }^{\mu }=2F^{\mu \lambda }F_{\nu \lambda }-\frac{1}{2}\,\delta _{\nu
}^{\mu }\,F^{2}+\mathcal{O}\left( \frac{1}{b^{2}}\right) \,,
\end{equation}
such that, as we mentioned above, the limit $b\rightarrow \infty $
reproduces the stress tensor for Maxwell electrodynamics.

Using the Stokes' theorem, the surface term in the variation (\ref{var_Ibulk})
can be written as
\begin{equation}
\Theta =\frac{1}{16\pi G}\,\sqrt{-h}\,n_{\mu }\,\left( \delta _{\lbrack
\alpha \beta ]}^{[\mu \nu ]}\,g^{\alpha \gamma }\delta \hat{\Gamma}_{\gamma
\nu }^{\beta }+\frac{4F^{\mu \nu }\delta A_{\nu }}{\sqrt{1+\frac{F^{2}}{%
2b^{2}}}}\right) +c_{d}\,\delta B_{d}\,.  \label{theta}
\end{equation}
We consider a manifold with a single boundary at radial infinity such that
we can choose a radial foliation in Gaussian form,
\begin{equation}
ds^{2}=g_{\mu \nu }\,dx^{\mu }dx^{\nu }=N^{2}\left( r\right)
\,dr^{2}+h_{ij}(r,x)\,dx^{i}dx^{j}\,,\qquad \qquad \sqrt{-g}=N\sqrt{-h}\,.
\label{radial foliation}
\end{equation}
Here $h_{ij}$ describes the intrinsic geometry of the spacelike boundary
$\partial M$ parameterized by the coordinate set $\left\{ x^{i}\right\} $,
whereas its extrinsic properties are given in terms of the outward-pointing
unit normal $n_{\mu }=\left( n_{r},n_{i}\right) =(N,\vec{0})$. When one
projects the fully-covariant expression $\Theta $ to the boundary, one risks
that there would be components of the Christoffel symbol that are not
expressible as tensorial quantities on $\partial M$. However, the only
relevant components of the connection $\hat{\Gamma}_{\mu \nu }^{\alpha }$
are indeed expressed in terms of the extrinsic curvature
$K_{ij}=-\frac{1}{2N}\,h_{ij}^{\prime }$ as
\begin{equation}
\hat{\Gamma}_{ij}^{r}=\frac{1}{N}\,K_{ij\,},\qquad
\hat{\Gamma}_{rj}^{i}=-NK_{j}^{i}\,,\qquad \hat{\Gamma}_{rr}^{r}
=\frac{N^{\prime }}{N}\,,
\label{KChr}
\end{equation}
where the prime denotes radial derivative. Thus, the surface term
(\ref{theta}) projected to the boundary takes the form
\begin{equation}
\Theta =-\frac{1}{16\pi G}\,\sqrt{-h}\,\left( \left( h^{-1}\delta h\right)
_{j}^{i}\,K_{i}^{j}+2\delta K_{i}^{i}-\frac{4NF^{ri}\delta A_{i}}{\sqrt{1+
\frac{F^{2}}{2b^{2}}}}\right) +c_{d}\,\delta B_{d}\,.
\end{equation}

The radial foliation (\ref{radial foliation}) implies the Gauss-Codazzi
relations for the spacetime curvature, as well,
\begin{eqnarray}
\hat{R}_{kl}^{ir} &=&\frac{1}{N}\,\left( \nabla _{l}K_{k}^{i}-\nabla
_{k}K_{l}^{i}\right) \,,  \label{Codazzi2} \\
\hat{R}_{kr}^{ir} &=&\frac{1}{N}\,\left( K_{k}^{i}\right) ^{\prime
}-K_{l}^{i}\,K_{k}^{l}\,,  \label{Codazzi3} \\
\hat{R}_{kl}^{ij}
&=&R_{kl}^{ij}(h)-K_{k}^{i}\,K_{l}^{j}+K_{l}^{i}\,K_{k}^{j}\,\equiv
R_{kl}^{ij}-K_{[k}^{[i}K_{l]}^{j]}\,,  \label{Codazzi1}
\end{eqnarray}
where $\nabla _{l}=\nabla _{l}(\Gamma _{ij}^{k})$ is the covariant
derivative defined in the Christoffel symbol of the boundary
$\hat{\Gamma}_{ij}^{k}(g)=\Gamma _{ij}^{k}(h)$.


\section{Topological \textbf{black hole solutions}}

A static black hole ansatz for the metric $g_{\mu \nu }$ in the coordinate
set $x^{\mu }=(t,r,\varphi ^{m})$ is given by the line element%
\begin{equation}
ds^{2}=-f(r)\,dt^{2}+\frac{dr^{2}}{f(r)}+r^{2}\gamma _{mn}(\varphi
)\,d\varphi ^{m}d\varphi ^{n}\,,\qquad \sqrt{-h}=r^{D-2}\sqrt{f}\,\sqrt{%
\gamma }\,.  \label{BH}
\end{equation}%
Here $\gamma _{mn}$ is the metric of a $(D-2)$-dimensional Riemann space $%
\Sigma _{D-2}$ with curvature $R_{m_{2}n_{2}}^{m_{1}n_{1}}(\gamma
)=k\,\delta _{\lbrack m_{2}n_{2}]}^{[m_{1}n_{1}]}$, so that $k=+1$, $0$\ or $%
-1$\ describes a spherical, locally flat or hyperbolic transversal section,
respectively. As a consequence, the horizon $r=r_{+}$ possess the same
topology, and it is defined by the largest root of the equation $f(r_{+})=0$.

The extrinsic curvature of the boundary has the simple form%
\begin{equation}
K_{j}^{i}=-\frac{1}{2}\,\sqrt{f}\,h^{ik}h_{kj}^{\prime }=\left(
\begin{array}{cc}
-\frac{f^{\prime }}{2\sqrt{f}} & 0 \\
0 & -\frac{\sqrt{f}}{r}\,\delta _{n}^{m}%
\end{array}%
\right) \,.  \label{K}
\end{equation}%
The non-vanishing components of the\ intrinsic curvature are%
\begin{equation}
R_{m_{2}n_{2}}^{m_{1}n_{1}}(h)=\frac{k}{r^{2}}\,\delta _{\lbrack
m_{2}n_{2}]}^{[m_{1}n_{1}]}\,.  \label{R(h)}
\end{equation}

We choose a gauge field with dependence on the radial coordinate%
\begin{equation}
A_{\mu }=\phi \left( r\right) \,\delta _{\mu }^{t}\,,  \label{A}
\end{equation}%
with the associated field strength%
\begin{equation}
F_{\mu \nu }=E(r)\,\delta _{\lbrack \mu \nu ]}^{[tr]}\,,  \label{F}
\end{equation}%
where the electric field is given by
\begin{equation}
E(r)=-\phi ^{\prime }(r)\,.  \label{def_E}
\end{equation}%
The equation of motion $\mathcal{E}^{\mu }=0$ determines the form of the
electric field as%
\begin{equation}
E(r)=\frac{q}{\sqrt{\frac{q^{2}}{b^{2}}+r^{2D-4}}}\,,  \label{E(r)}
\end{equation}%
where $q$ is an integration constant related to the electric charge, since $%
E $ behaves as $\frac{q}{r^{D-2}}+\mathcal{O}\left( \frac{1}{b^{2}}\right) $%
. The $U(1)$ gauge potential measured with respect to the horizon is%
\begin{equation}
\phi (r)=-\int\limits_{r_{+}}^{r}dv\,E(v)\,,  \label{integral_phi}
\end{equation}%
that, for EBI gravity, can be written down in terms of a hypergeometric
function $_{2}F_{1}\left( a,b;c;z\right) $ (see Appendix \ref{hyper}) as
\cite{Dehghani-Sedehi}%
\begin{equation}
\phi (r)=\frac{q}{\left( D-3\right) r^{D-3}}\,_{2}F_{1}\left( \frac{1}{2},%
\frac{D-3}{2D-4};\frac{3D-7}{2D-4};-\frac{q^{2}}{b^{2}r^{2D-4}}\right) -\Phi
\,.  \label{solution_phi}
\end{equation}%
The integration constant $\Phi $ is fixed by the condition $\phi (r_{+})=0$,
that is,%
\begin{equation}
\Phi =\frac{q}{\left( D-3\right) r_{+}^{D-3}}\,_{2}F_{1}\left( \frac{1}{2},%
\frac{D-3}{2D-4};\frac{3D-7}{2D-4};-\frac{q^{2}}{b^{2}r_{+}^{2D-4}}\right)
\,.  \label{phi}
\end{equation}%
Note that the asymptotic value of the potential is $\phi (\infty )=-\Phi $.
As we will see below, the constant $\Phi $ plays an important role in black
hole thermodynamics as it is the conjugated variable to the $U(1)$ charge
for Born-Infeld theory.

A useful property of the hypergeometric function in (\ref{phi}) for later
purposes is
\begin{equation}
\frac{d}{dr}\left( \frac{_{2}F_{1}}{\left( D-3\right) r^{D-3}}\right) =-%
\frac{1}{\sqrt{\frac{q^{2}}{b^{2}}+r^{2D-4}}}\,.  \label{d_hyper}
\end{equation}%
The equation of motion $\mathcal{E}_{t}^{t}=0\ $in the black hole ansatz (%
\ref{BH}), with the field strength as given by Eq. (\ref{F}), adopts the form%
\begin{equation}
rf^{\prime }+\left( D-3\right) \left( f-k\right) -\left( D-1\right) \,\frac{%
r^{2}}{\ell ^{2}}-\frac{4b^{2}r^{2}}{D-2}\left( 1-\frac{1}{\sqrt{1-\frac{%
E^{2}}{b^{2}}}}\right) =0\,.  \label{diff_f}
\end{equation}%
Plugging in the explicit form of the electric field (\ref{E(r)}) into the
above relation, the solution for the metric function is \cite%
{Fernando-Krug,Dey,Cai-Pang-Wang}%
\begin{eqnarray}
f(r) &=&k+\frac{r^{2}}{\ell ^{2}}-\frac{\mu }{r^{D-3}}+\frac{4b^{2}r^{2}}{%
\left( D-1\right) \left( D-2\right) }\left( 1-\sqrt{1+\frac{q^{2}}{%
b^{2}r^{2D-4}}}\right)  \notag \\
&&+\frac{4q^{2}}{\left( D-1\right) \left( D-3\right) r^{2D-6}}%
\,_{2}F_{1}\left( \frac{1}{2},\frac{D-3}{2D-4};\frac{3D-7}{2D-4};-\frac{q^{2}%
}{b^{2}r^{2D-4}}\right) \,,  \label{f}
\end{eqnarray}%
where $\mu \,$is a parameter of dimension [mass]$\times $[Newton's
constant]. The remaining equations of motion are identically satisfied for\
the function $f(r)$\ given above. Then, the horizon radius $r_{+}$ can be
obtained from the equation%
\begin{equation}
0=f(r_{+})=k+\frac{r_{+}^{2}}{\ell ^{2}}-\frac{\mu }{r_{+}^{D-3}}+\frac{%
4b^{2}r_{+}^{2}}{\left( D-1\right) \left( D-2\right) }\left( 1-\sqrt{1+\frac{%
q^{2}}{b^{2}r_{+}^{2D-4}}}\right) +\frac{4q\Phi }{\left( D-1\right)
r_{+}^{D-3}}\,.  \label{horizon}
\end{equation}

The expansion of $f(r)$ goes as%
\begin{equation}
f(r)=k+\frac{r^{2}}{\ell ^{2}}-\frac{\mu }{r^{D-3}}+\frac{2q^{2}}{\left(
D-2\right) \left( D-3\right) r^{2D-6}}+\mathcal{O}\left( \frac{r^{-4D+10}}{%
b^{2}}\right) \,,  \label{f_Maxwell}
\end{equation}%
which, in the Maxwell's limit, reduces to the metric function of the
Reissner-Nordstr\"{o}m-AdS black hole (see, e.g.,
\cite{Chamblin-Emparan-Johnson-Myers}) and, for $q=0$, to the one of
topological Schwarzschild-AdS black holes.
Extreme black hole has a
degenerate horizon at $r_{ext}$, when both $f(r_{ext})$ and
$f^{\prime }(r_{ext})$ vanish. The mass for the
extreme solution in terms of the extremal charge $q_{ext}$ and the radius $%
r_{ext}$ is given by
\begin{equation}
\mu _{ext}=\frac{2k\,r_{ext}^{D-3}}{D-1}+\frac{4q_{ext}^{2}}{\left(
D-1\right)
\left( D-3\right) r_{ext}^{D-3}}\,_{2}F_{1}\left( \frac{1}{2},\frac{D-3}{2D-4%
};\frac{3D-7}{2D-4};-\frac{q_{ext}^{2}}{b^{2}r_{ext}^{2D-4}}\right)
\,, \label{mass ext}
\end{equation}%
and, in turn, the extremal charge is given by
\begin{equation}
q_{ext}^{2}=\frac{1}{4}\,\left( D-2\right) r_{ext}^{2D-4}\left(
\frac{\left(
D-3\right) k}{r_{ext}^{2}}+\frac{D-1}{\ell ^{2}}\right) \left[ 2+\frac{D-2}{%
4b^{2}}\left( \frac{\left( D-3\right) k}{r_{ext}^{2}}+\frac{D-1}{\ell ^{2}}%
\right) \right] \,,  \label{qextremal}
\end{equation}%
such that the solution is characterized by a single parameter.

It can be easily proved that extremality condition for the
Reissner-Nordstr\"{o}m-AdS black holes

\begin{equation}
\mu _{ext}^{2}=\frac{8}{\left( D-2\right) \left( D-3\right)
}\,q_{ext}^{2}\, \frac{\left( k+\frac{\left( D-2\right)
 r_{ext}^{2}}{\left( D-3\right) \ell ^{2}}
 \right) ^{2}}{\left( k+\frac{\left( D-1\right) r_{ext}^{2}}{\left(
D-3\right) \ell ^{2}}\right) }\,, \label{mass ext RN}
\end{equation}
 is recovered combining Eqs.(\ref{mass ext}) and (\ref{qextremal}), and taking the suitable limit.
 In the latter relation, the vanishing cosmological constant limit (consistent only with $k=+1$) leads to
 an extremal mass proportional to the electric charge.

 A solution
with mass parameter $\mu<\mu_{ext}$ corresponds to a naked
singularity. Depending on the parameters $\mu $ and $q$,
Born-Infeld-AdS solutions with $\mu>\mu_{ext}$ may have two horizons
(so-called RN-AdS type black hole) or one horizon (Schwarzschild-AdS
type black hole). Thus, in general, a diagram $(q,\mu)$ contains the
three types of solutions, which meet at the \emph{triple point},
where all phases coexist \cite{Cai-Pang-Wang,Fernando,Dey}. This
behavior resembles the one in liquid-gas-solid phase diagram, for
RNAdS type, naked singularities and Schwarzschild-AdS type
solutions, respectively.
 The nonlinear nature of the
theory makes possible this richer phase structure, which does not
appear in Reissner-Nordstr\"{o}m-AdS case.

The interpretation of the parameters $\mu $ and $q$ in the general
solution in terms of conservation laws associated to global
symmetries confronts us with the long-standing problem of definition
of conserved charges in AdS gravity. In the context of AdS/CFT
correspondence, a finite, background-independent expression for the
conserved quantities is achieved through the addition of standard
counterterms to the bulk action. However, in this method, it is not
possible to get a closed formula for the asymptotic charges in an
arbitrary
dimension, as the counterterms series itself has not been found yet for any $%
D$.

In what follows, we show that the Kounterterms series is a suitable
prescription to deal with a regularization problem of both conserved
quantities and Euclidean action for EBI AdS black holes.

\section{Black hole thermodynamics in even dimensions}

\subsection{Variational principle and regularization problem}

Let us consider the pure AdS gravity action in four dimensions supplemented
by the Gauss-Bonnet term

\begin{equation}
I_{4}=-\int\limits_{M}d^{4}x\,\sqrt{-g}\left[ \frac{1}{16\pi G}\,\left( \hat{%
R}-2\Lambda \right) +\alpha \,(\hat{R}^{\mu \nu \sigma \rho }\hat{R}_{\mu
\nu \sigma \rho }-4\hat{R}^{\mu \nu }\hat{R}_{\mu \nu }+\hat{R}^{2})\right] ,
\end{equation}%
where $\alpha $\ is an arbitrary coupling constant. Because the
Euler-Gauss-Bonnet term in four dimensions is a topological invariant, the
bulk dynamics is not modified by this combination of quadratic-curvature
terms.

It has been claimed that this freedom might shift the value of the entropy
by a constant proportional to $\alpha $, as a direct application of the
Wald's formalism \cite{Wald}. However, an arbitrary coupling of the Euler
term is inconsistent from the point of view of the action principle. In
fact, the theory has a well-posed variational principle only if the coupling
constant is properly adjusted as $\alpha =\ell ^{2}/(64\pi G)$,\ as shown in
Ref.\cite{Aros-Contreras-Olea-Troncoso-Zanelli-4D}. This reasoning results
in the on-shell cancellation of divergences in the conserved quantities.

For an arbitrary $\alpha $, the Gauss-Bonnet invariant introduces
additional divergent terms in the Euclidean action. It is,
therefore, remarkable that fixing its coupling constant as above,
also provides a mechanism to regularize the Euclidean action for
asymptotically AdS spacetimes and to reproduce the correct black
hole thermodynamics \cite{Olea-KerrBH}. Nevertheless, the entropy is
still modified by an additive constant. In case of static black
holes, this constant is given by $ \ell ^{2} V(\Sigma _{2})\chi
(\Sigma _{2})/8G$, where $V (\Sigma _{2})$\ stands for the volume of
the two-dimensional transversal section, which has an Euler
characteristic $\chi (\Sigma _{2})=2k$. Unfortunately, for
topological Schwarzschild-AdS solutions with $k=-1$, this value
leads to negative entropy for black holes with $r_{+}<\ell $, what
is clearly a drawback.

The way to avoid this inconsistency is to consider, instead, the closed form
that is locally equivalent to the Gauss-Bonnet term, i.e., the second Chern
form. Since for a given spacetime foliation this boundary term is written in
terms of the extrinsic and intrinsic curvatures of the boundary, what we
have is an alternative counterterm series, that does not depend only on
intrinsic tensors on $\partial M$ \cite{Olea-KerrBH}.

In higher even dimensions ($D=2n$), the situation is quite similar, only the
Euler term is not longer quadratic in the curvature
\cite{Aros-Contreras-Olea-Troncoso-Zanelli-2n}. Taking then the $n$-th Chern form
\begin{eqnarray}
B_{2n-1} & = &  2n\sqrt{-h}\int\limits_{0}^{1}dt\,\delta _{\lbrack i_{1}\cdots
i_{2n-1}]}^{[j_{1}\cdots j_{2n-1}]}\,K_{j_{1}}^{i_{1}}\times \notag \\
&&\qquad \times
\left( \frac{1}{2}
\,R_{j_{2}j_{3}}^{i_{2}i_{3}}-t^{2}K_{j_{2}}^{i_{2}}K_{j_{3}}^{i_{3}}\right)
\cdots \left( \frac{1}{2}
\,R_{j_{2n-2}j_{2n-1}}^{i_{2n-2}i_{2n-1}}-t^{2}K_{j_{2n-2}}^{i_{2n-2}}K_{j_{2n-1}}^{i_{2n-1}}\right) \,,
\label{B-even}
\end{eqnarray}
the action (\ref{I}) becomes finite if we adjust the coefficient of this
boundary term as
\begin{equation}
c_{2n-1}=\frac{1}{16\pi G}\frac{\left( -1\right) ^{n}\ell ^{2n-2}}{n\left(
2n-2\right) }\,.
\end{equation}
One can also understand the Kounterterms for this case as the correction to
the Euler characteristic due to the boundary in any even-dimensional
manifold (a transgression form for the Lorentz group $SO(2n-1,1)$
\cite{Eguchi-Gilkey-Hanson}). Using the boundary formulation locally equivalent
to bulk topological invariants would make easier a possible comparison with
the standard regularization procedure.

The term (\ref{B-even}) varies (up to a total derivative) as
\begin{eqnarray}
\delta B_{2n-1} &=&\frac{n}{2^{n-1}}\sqrt{-h}\,\delta _{\lbrack i_{1}\cdots
i_{2n-1}]}^{[j_{1}\cdots j_{2n-1}]}\,\left[ \left( h^{-1}\delta h\right)
_{k}^{i_{1}}K_{j_{1}}^{k}+2\delta K_{j_{1}}^{i_{1}}\right] \times  \notag \\
&&\qquad \times \left(
R_{j_{2}j_{3}}^{i_{2}i_{3}}-K_{[j_{2}}^{[i_{2}}K_{j_{3}]}^{i_{3}]}\right)
\cdots \left(
R_{j_{2n-2}j_{2n-1}}^{i_{2n-2}i_{2n-1}}-K_{[j_{2n-2}}^{[i_{2n-2}}K_{j_{2n-1}]}^{i_{2n-1}]}\right) \,,
\end{eqnarray}
so that the complete action (\ref{I}) on-shell changes under arbitrary
variations in the following way,
\begin{eqnarray}
\delta I_{2n} &=&\frac{nc_{2n-1}}{2^{n-1}}\int\limits_{\partial
M}d^{2n-1}x\,\sqrt{-h}\,\delta _{\lbrack i_{1}\cdots
i_{2n-1}]}^{[j_{1}\cdots j_{2n-1}]}\left[ \left( h^{-1}\delta
h\right)
_{k}^{i_{1}}K_{j_{1}}^{k}+2\delta K_{j_{1}}^{i_{1}}\right] \times  \notag \\
&&\times \left[ \left(
\,R_{j_{2}j_{3}}^{i_{2}i_{3}}-K_{[j_{2}}^{[i_{2}}K_{j_{3}]}^{i_{3}]}\right)
\cdots \left(
\,R_{j_{2n-2}j_{2n-1}}^{i_{2n-2}i_{2n-1}}
-K_{[j_{2n-2}}^{[i_{2n-2}}K_{j_{2n-1}]}^{i_{2n-1}]}\right)- \right. \notag\\
&& -\left.
\frac{(-1)^{n-1}}{\ell ^{2(n-1)}}\,\delta _{\lbrack
j_{2}j_{3}]}^{[i_{2}i_{3}]}\cdots \delta _{\lbrack
j_{2n-2}j_{2n-1}]}^{[i_{2n-2}i_{2n-1}]}\right] +\frac{1}{4\pi G}
\int\limits_{\partial M}d^{2n-1}x\,\sqrt{-h}\,
\frac{NF^{ri}\delta A_{i}}{\sqrt{1+\frac{F^{2}}{2b^{2}}}}\,.
\label{varI_2n}
\end{eqnarray}
Using the Gauss-Codazzi equation (\ref{Codazzi1}), it is easy to prove that
the second line in the above expression can be factorized by
$\hat{R}_{kl}^{ij}+\frac{1}{\ell ^{2}}\,\delta _{\lbrack kl]}^{[ij]}$, so that
\begin{eqnarray}
\delta I_{2n} &=&\frac{n\left( n-1\right) c_{2n-1}}{2^{n-1}}
\int\limits_{\partial M}\int\limits_{0}^{1}dt\,\delta _{\lbrack
i_{1}\cdots i_{2n-1}]}^{[j_{1}\cdots j_{2n-1}]}\left[ \left(
h^{-1}\delta h\right) _{k}^{i_{1}}K_{j_{1}}^{k}+2\delta
K_{j_{1}}^{i_{1}}\right] \,\left(
\hat{R}_{j_{2}j_{3}}^{i_{2}i_{3}}+\frac{1}{\ell ^{2}}\,\delta
_{\lbrack
j_{2}j_{3}]}^{[i_{2}i_{3}]}\right) \times  \notag \\
&&\times \left[ t\left( \hat{R}_{j_{4}j_{5}}^{i_{4}i_{5}}+\frac{1}{\ell ^{2}}
\,\delta _{\lbrack j_{4}j_{5}]}^{[i_{4}i_{5}]}\right) \cdots \left(
\hat{R}_{j_{2n-2}j_{2n-1}}^{i_{2n-2}i_{2n-1}}+\frac{1}{\ell ^{2}}\,\delta _{\lbrack
j_{2n-2}j_{2n-1}]}^{[i_{2n-2}i_{2n-1}]}\right)- \right. \notag \\
&&\qquad\left.-\frac{1}{\ell ^{2(n-1)}}
\,\delta _{\lbrack j_{4}j_{5}]}^{[i_{4}i_{5}]}\cdots \delta _{\lbrack
j_{2n-2}j_{2n-1}]}^{[i_{2n-2}i_{2n-1}]}\right]
 +\frac{1}{4\pi G}\int\limits_{\partial M}d^{2n-1}x\,\sqrt{-h}\,
\frac{NF^{ri}\delta A_{i}}{\sqrt{1+\frac{F^{2}}{2b^{2}}}}\,.
\end{eqnarray}
Thus, in order to ensure that the action is stationary under arbitrary
variations, we take the condition on the curvature
\begin{equation}
\hat{R}_{\alpha \beta }^{\mu \nu }+\frac{1}{\ell ^{2}}\,\delta _{\lbrack
\alpha \beta ]}^{[\mu \nu ]}=0\,,\qquad \text{at }\partial M\,,  \label{bc R}
\end{equation}
that, in particular, is valid for the boundary indices. The rest of the
surface term is cancelled assuming that the transversal components of the
gauge field satisfy
\begin{equation}
\delta A_{i}=0\,,\qquad \text{at }\partial M\,.  \label{bc A}
\end{equation}
The condition (\ref{bc R}) simply means that the spacetime has constant
curvature in the asymptotic region (asymptotically locally AdS).


\subsection{Conserved quantities \label{conserved}}

The spacetime diffeomorphisms $\delta x^{\mu }=\xi ^{\mu }(x)$ generate the
changes of the dynamical fields $g_{\mu \nu }$ and $A_{\mu }$, whose
infinitesimal form is given in terms of the Lie derivative,
\begin{eqnarray}
\delta _{\xi }g_{\mu \nu } &=&\pounds _{\xi }g_{\mu \nu }\equiv -\left( \hat{
\nabla}_{\mu }\xi _{\nu }+\hat{\nabla}_{\nu }\xi _{\mu }\right) \,, \\
\delta _{\xi }A_{\mu } &=&\pounds _{\xi }A_{\mu }\equiv -\partial _{\mu
}\left( \xi ^{\nu }A_{\nu }\right) +\xi ^{\nu }F_{\mu \nu }\,.
\end{eqnarray}
Then, the EBI action (\ref{I}) transforms under the diffeomorphisms
according to
\begin{eqnarray}
\delta _{\xi }I &=&\int\limits_{M}d^{D}x\,\left[ \pounds _{\xi
}\left( \sqrt{-g}\,\mathcal{L}\right) +\partial _{\mu }\left(
\sqrt{-g}\,\xi ^{\mu }\,\mathcal{L}\right) \right]
+c_{d}\,\int\limits_{\partial M}d^{d}x\,\left[ \pounds _{\xi
}B_{d}+\partial _{i}\left( \xi ^{i}B_{d}\right) \right]  \notag
\\
&=&\int\limits_{\partial M}d^{d}x\,\sqrt{-h}\,n_{\mu }\left(
\Theta ^{\mu }(\xi )+\xi ^{\mu
}\mathcal{L}+\frac{1}{\sqrt{-h}}\,c_{d}\,n^{\mu
}\,\partial _{i}\left( \xi ^{i}B_{d}\right) \right) +\text{e.o.m.\thinspace }
,  \label{diff_Ibulk}
\end{eqnarray}
where $\Theta =\sqrt{-h}\,n_{\mu }\Theta ^{\mu }$. It has been used that the
volume element, Jacobian and Lagrangian density transform as
\begin{eqnarray}
\delta _{\xi }\left( d^{D}x\right) &=&\partial _{\mu }\xi ^{\mu }\,d^{D}x\,,
\\
\pounds _{\xi }\sqrt{-g} &=&-\sqrt{-g}\,\hat{\nabla}_{\mu }\xi ^{\mu }\,, \\
\delta _{\xi }\mathcal{L} &=&\pounds _{\xi }\mathcal{L}+\xi ^{\mu }\partial
_{\mu }\mathcal{L\,},
\end{eqnarray}
respectively. The Lie derivative acts on the Lagrangian as $\pounds _{\xi }L=
\frac{\partial \mathcal{L}}{\partial g_{\mu \nu }}\,\pounds _{\xi }g_{\mu
\nu }+\frac{\partial \mathcal{L}}{\partial \hat{\Gamma}_{\mu \nu }^{\beta }}
\,\pounds _{\xi }\hat{\Gamma}_{\mu \nu }^{\beta }+\frac{\partial \mathcal{L}
}{\partial A_{\mu }}\,\pounds _{\xi }A_{\mu }$. Similar expressions hold for
the boundary term as well.

The diffeormorphic invariance ($\delta _{\xi }I=0$) defines the Noether
current%
\begin{equation}
J^{\mu }=\Theta ^{\mu }(\xi )+\xi ^{\mu }\mathcal{L}+\frac{1}{\sqrt{-h}}
\,c_{d}\,n^{\mu }\,\partial _{i}\left( \xi ^{i}B_{d}\right) \mathcal{\,}.
\label{J^mu}
\end{equation}

Using (\ref{theta}), we write out the surface term as
\begin{equation}
\Theta (\xi )=\frac{1}{16\pi G}\,\sqrt{-h}\,n_{\mu }\left( \delta _{\lbrack
\alpha \beta ]}^{[\mu \nu ]}\,g^{\alpha \gamma }\pounds _{\xi }
\hat{\Gamma}_{\gamma \nu }^{\beta }
+\frac{4F^{\mu \nu }\pounds _{\xi }A_{\nu }}{\sqrt{1+
\frac{F^{2}}{2b^{2}}}}\right) +c_{d}\,\pounds _{\xi }B_{d}\,.
\label{surfLie}
\end{equation}
The fact that the current (\ref{J^mu}) satisfies $\partial _{\mu }\left(
\sqrt{-g}\,J^{\mu }\right) =0$ means, by virtue of the Poincar\'{e}'s lemma,
that it can always be written locally as an exact form. However, only when
the current can be cast as a total derivative globally at the boundary, the
Noether charge can be directly read off from it. The radial foliation
(\ref{radial foliation}) defines a conservation law along the radial cooordinate,
such that the quantity $Q[\xi ]=\int_{\partial M}\sqrt{-g}\,J^{r}$ is a
constant of motion.

We take a timelike ADM foliation for the line element on $\partial
M$ with the coordinates $x^{i}=\left( t,y^{m}\right) $, as
\begin{equation}
h_{ij}\,dx^{i}dx^{j}=-\tilde{N}^{2}(t)dt^{2}+\sigma _{mn}(dy^{m}
+\tilde{N}^{m}dt)(dy^{n}+\tilde{N}^{n}dt)\,,\qquad \sqrt{-h}=\tilde{N}\sqrt{\sigma }\,,
\end{equation}
that is generated by the outward-pointing unit normal vector
$u_{i}=(u_{t},u_{m})=(-\tilde{N},\vec{0})$. $\sigma _{mn}$ represents the
metric of the boundary of spatial section at constant time
$\Sigma _{\infty} $.

If the radial component of the current adopts the form
\begin{equation}
\sqrt{-g}\,J^{r}=\partial _{j}(\sqrt{-h}\,\xi ^{i}\,q_{i}^{j})\,,
\end{equation}
the Noether theorem provides the conserved charges $Q[\xi ]$ of the theory
as surface integrals on $\Sigma _{\infty }$ as
\begin{equation}
Q[\xi ]=\int\limits_{\Sigma _{\infty }}d^{2n-2}y\,\sqrt{\sigma }
\,u_{j}\,\xi ^{i}\,q_{i}^{j}\,,
\end{equation}
for a given set of asymptotic Killing vectors $\{\xi \}$.

In even dimensions, the expression for $\Theta (\xi )$ is obtained from
(\ref{varI_2n}) by suitable projection on the boundary as
\begin{eqnarray}
n_{\mu }\Theta ^{\mu }(\xi ) &=&\frac{nc_{2n-1}}{2^{n-1}}\,\,\delta
_{\lbrack i_{1}\cdots i_{2n-1}]}^{[j_{1}\cdots j_{2n-1}]}\left[
\left( h^{-1}\pounds _{\xi }h\right) _{k}^{i_{1}}K_{j_{1}}^{k}+2\pounds _{\xi
}K_{j_{1}}^{i_{1}}\right] \times   \notag \\
&&\times \left[ \left(
R_{j_{2}j_{3}}^{i_{2}i_{3}}-K_{[j_{2}}^{[i_{2}}K_{j_{3}]}^{i_{3}]}\right)
\cdots \left(
R_{j_{2n-2}j_{2n-1}}^{i_{2n-2}i_{2n-1}}
-K_{[j_{2n-2}}^{[i_{2n-2}}K_{j_{2n-1}]}^{i_{2n-1}]}\right) - \right. \notag \\
&&-\left.\frac{(-1)^{n-1}}{\ell ^{2(n-1)}}\,\delta _{\lbrack
j_{2}j_{3}]}^{[i_{2}i_{3}]}\cdots \delta _{\lbrack
j_{2n-2}j_{2n-1}]}^{[i_{2n-2}i_{2n-1}]}\right]
+\frac{1}{16\pi G}\,\frac{4NF^{ri}\pounds _{\xi }A_{i}}{\sqrt{1+
\frac{F^{2}}{2b^{2}}}}\,.
\end{eqnarray}
Using Eq.(\ref{KChr}), the action of the diffeomorphism on the extrinsic
curvature can be worked out from the corresponding components of the
Christoffel symbol
\begin{equation}
\pounds _{\xi }\hat{\Gamma}_{\mu \nu }^{\alpha }=\frac{1}{2}\,\left(
\hat{R}_{\ \mu \nu \beta }^{\alpha }+\hat{R}_{\ \nu \mu \beta }^{\alpha }\right)
\xi ^{\beta }-\frac{1}{2}\,\left( \hat{\nabla}_{\mu }\hat{\nabla}_{\nu }\xi
^{\alpha }+\hat{\nabla}_{\nu }\hat{\nabla}_{\mu }\xi ^{\alpha }\right) .
\end{equation}
The procedure defines the Noether charge as
\begin{eqnarray}
Q\left[ \xi \right]
&=&\frac{nc_{2n-1}}{2^{n-2}}\int\limits_{\Sigma _{\infty
}}d^{2n-2}y\,\sqrt{\sigma }\,u_{j}\,\xi ^{i}\delta _{\lbrack
i_{1}i_{2}\cdots i_{2n-1}]}^{[jj_{2}\cdots
j_{2n-1}]}\,K_{i}^{i_{1}}\times
\notag \\
&&\qquad \qquad \times \left( \hat{R}_{j_{2}j_{3}}^{i_{2}i_{3}}\cdots
\hat{R}_{j_{2n-2}j_{2n-1}}^{i_{2n-2}i_{2n-1}}-\frac{\left( -1\right) ^{n-1}}{\ell
^{2(n-1)}}\,\delta _{\lbrack j_{2}j_{3}]}^{[i_{2}i_{3}]}\cdots \delta
_{\lbrack j_{2n-2}j_{2n-1}]}^{[i_{2n-2}i_{2n-1}]}\right) \,.  \label{Q(2n)}
\end{eqnarray}
The second line in the above expression can be always factorized by $\left(
\hat{R}_{j_{2}j_{3}}^{i_{2}i_{3}}+\frac{1}{\ell ^{2}}\,\delta _{\lbrack
j_{2}j_{3}]}^{[i_{2}i_{3}]}\right) $,\ that means that the charge is
identically zero for any global constant-curvature spacetime. The mass of
the EBI AdS black hole solution (\ref{BH}, \ref{f}) is computed evaluating
the formula (\ref{Q(2n)}) for the Killing vector $\xi ^{i}=\left( 1,\vec{0}
\right) $,
\begin{eqnarray}
Q\left[ \partial _{t}\right]  &=&-\frac{nc_{2n-1}\,}{2^{n-2}}
\int\limits_{\Sigma _{2n-2}}d^{2n-2}\varphi \sqrt{\gamma }\sqrt{f}
\,r^{2n-2}\,\delta _{\lbrack n_{1}\cdots n_{2n-2}]}^{[m_{1}\cdots
m_{2n-2}]}\,K_{t}^{t}\times   \notag \\
&&\qquad \qquad \times \left( \hat{R}_{m_{1}m_{2}}^{n_{1}n_{2}}\cdots
\hat{R}_{m_{2n-3}m_{2n-2}}^{n_{2n-3}n_{2n-2}}-\frac{(-1)^{n-1}}{\ell ^{2(n-1)}}
\,\delta _{\lbrack m_{1}m_{2}]}^{[n_{1}n_{2}]}\cdots \delta _{\lbrack
m_{2n-3}m_{2n-2}]}^{[n_{2n-3}n_{2n-2}]}\right) \,,
\end{eqnarray}
and we get
\begin{equation}
Q\left[ \partial _{t}\right] =\frac{V(\Sigma _{2n-2})}{16\pi G}
\,\lim_{r\rightarrow \infty }\,f^{\prime }\left[ r^{2n-2}-\ell ^{2n-2}\left(
f-k\right) ^{n-1}\right] \,.  \label{Noether_2n}
\end{equation}
With the help of the identity $a^{n-1}-b^{n-1}=\left( a-b\right)
a^{n-2}\sum_{p=0}^{n-2}\left( \frac{b}{a}\right) ^{p}$ and using the
asymptotic expansions
\begin{eqnarray}
r^{2n-3}\left( f-k-\frac{r^{2}}{\ell ^{2}}\right)  &=&-\mu +\mathcal{O}
\left( \frac{1}{r^{2n-3}}\right) \,, \\
\frac{1}{r}\,f^{\prime } &=&\frac{2}{\ell ^{2}}+\mathcal{O}\left(
\frac{1}{r^{2n-3}}\right) \,, \\
\frac{\ell ^{2}}{r^{2}}\,\left( f-k\right)  &=&1+\mathcal{O}\left(
\frac{1}{r^{2n-1}}\right) \,,
\end{eqnarray}
we see that all divergences at radial infinity are cancelled out, such that
the energy of the EBI AdS black hole in even dimensions is
\begin{equation}
\mathcal{E}=Q\left[ \partial _{t}\right] =\frac{\left( D-2\right) V(\Sigma
_{D-2})\,\mu }{16\pi G}\equiv M\,.  \label{M}
\end{equation}
It is straightforward to check that the contribution of the Born-Infeld
electromagnetic term to the charge in any dimension $D$ vanishes. In fact,
it can be worked out from the general form of the surface term
(\ref{surfLie}), that the extra piece in the charge is
\begin{equation}
Q_{BI}\left[ \xi \right] =\frac{1}{16\pi G}\int\limits_{\Sigma
_{\infty }}d^{2n-1}y\,\sqrt{\sigma }\,u_{j}\,\frac{4NF^{rj}\left(
\xi ^{i}A_{i}\right) }{\sqrt{1+\frac{F^{2}}{2b^{2}}}}\,.
\end{equation}
Using Eq.(\ref{limQ}) and asymptotic condition $\phi (\infty )=-\Phi $, we
find that
\begin{equation}
Q_{BI}\left[ \partial _{t}\right] =-Q\Phi \,\lim_{r\rightarrow \infty }
\frac{1}{\sqrt{f}}=0\,,
\end{equation}
since $\frac{1}{\sqrt{f}}$ behaves as $\mathcal{O}\left( \frac{1}{r}\right) $
at large distances.

Next we calculate a Noether charge associated to a $U(1)$ gauge
transformation $\delta _{\lambda }A_{\mu }=\partial _{\mu }\lambda $,
$\delta _{\lambda }g_{\mu \nu }=0$. The EBI action in any dimension $D$
changes by a boundary term,
\begin{equation}
\delta _{\lambda }I=\frac{1}{4\pi G}\int\limits_{\partial M}d^{D-1}x\,
\sqrt{-h}\,\frac{n_{\mu }F^{\mu \nu }\partial _{\nu }\lambda }{\sqrt{1
+\frac{F^{2}}{2b^{2}}}}\,,
\end{equation}
from where we can recognize a piece corresponding to the the equation of
motion
\begin{equation}
\sqrt{-h}\,\frac{n_{\mu }F^{\mu \nu }\partial _{\nu }\lambda }{\sqrt{1
+\frac{F^{2}}{2b^{2}}}}=\partial _{i}
\left( \frac{\sqrt{-h}\lambda NF^{ri}}{\sqrt{1+
\frac{F^{2}}{2b^{2}}}}\right) +\frac{1}{4}\,\lambda N\sqrt{-h}
\,\mathcal{E}^{r}\,,
\end{equation}
and project the total derivative on the surface $\Sigma _{\infty }$.
Assuming that $\lambda $ is constant at the boundary, the $U(1)$ charge is
\begin{equation}
Q=\frac{1}{4\pi G}\int\limits_{\Sigma _{\infty }}d^{D-1}y\sqrt{\sigma }
\,u_{i}\,\frac{NF^{ri}}{\sqrt{1+\frac{F^{2}}{2b^{2}}}}\,,
\end{equation}
that can also be written as
\begin{equation}
Q=\frac{V(\Sigma _{D-2})}{4\pi G}\lim_{r\rightarrow \infty }\,
\frac{r^{D-2}\,E}{\sqrt{1-\frac{E^{2}}{b^{2}}}}\,.  \label{limQ}
\end{equation}
Finally, evaluated for the black hole solution (\ref{BH}, \ref{f}), the
electric charge is
\begin{equation}
Q=\frac{V(\Sigma _{D-2})\,q}{4\pi G}\,.  \label{charge}
\end{equation}


\subsection{Regularized Euclidean action and Smarr relation}

The Euclidean continuation of the gravity action $I^{E}=-iI$ considers a
manifold that spans between the horizon $r_{+}$ and radial infinity. As the
horizon is shrunk to a point, the avoidance of a conical singularity at the
origin of the radial coordinate requires to identify the Euclidean time
$\tau =-it$ as $\tau \sim \tau +\beta $, where the period $\beta =\frac{1}{T}$
is the inverse of the Hawking temperature $T$,
\begin{equation}
T=\frac{1}{4\pi }\,\left. \frac{df(r)}{dr}\right\vert _{r=r_{+}}\,.
\label{HawkingT}
\end{equation}
We evaluate the temperature using Eq.(\ref{diff_f}) at the horizon, and we
obtain
\begin{equation}
T=\frac{1}{4\pi r_{+}}\,\left[ \left( D-3\right) k+\frac{\left( D-1\right)
r_{+}^{2}}{\ell ^{2}}+\frac{4b^{2}r_{+}^{2}}{D-2}\left( 1-\sqrt{1
+\frac{q^{2}}{b^{2}r_{+}^{2D-4}}}\right) \right] .  \label{T}
\end{equation}

In what follows, we work in the grand canonical ensemble, where the natural
variables are the temperature $T$ and the electric potential $\Phi $. The
Gibbs free energy $G(T,\Phi )=U-TS-\mathcal{Q}\Phi $, that satisfies the
differential equation $dG=-SdT-\mathcal{Q}d\Phi $, is represented by the
Euclidean action,
\begin{equation}
G=\frac{1}{\beta }\,I^{E}\,.  \label{Gibbs}
\end{equation}

The partition function in semiclassical approximation reads
\begin{equation}
Z=e^{-I^{E}}\,,
\end{equation}
from where the gravitational entropy is obtained as
\begin{equation}
S=\beta \,\left( \frac{\partial I^{E}}{\partial \beta }\right) _{\Phi
}-I^{E}\,,  \label{entropy}
\end{equation}
the internal energy is
\begin{equation}
U=\,\left( \frac{\partial I^{E}}{\partial \beta }\right) _{\Phi }-\frac{\Phi
}{\beta }\left( \frac{\partial I^{E}}{\partial \Phi }\right) _{\beta }\,,
\label{internal}
\end{equation}
and the thermodynamic charge is
\begin{equation}
\mathcal{Q}=-\frac{1}{\beta }\left( \frac{\partial I^{E}}{\partial \Phi }
\right) _{\beta }\,.  \label{TD_charge}
\end{equation}

First we calculate the bulk Euclidean action (\ref{EH-BI}) evaluated
for the black hole solution. Since the solution is static and
$\Sigma _{D-2}$ is a maximally symmetric submanifold, the
integrations along $\tau $ and $\varphi ^{m}$ are trivial, leading
to
\begin{eqnarray}
I_{bulk}^{E} &=&\frac{\beta V(\Sigma _{D-2})}{16\pi G}\int\limits_{r_{+}}^{%
\infty }dr\,r^{D-2}\,\left[ f^{\prime \prime }+\,2\left( D-2\right)
\,\frac{f^{\prime}}{r} +\left( D-2\right) \left(
D-3\right)  \dfrac{\left( f-k\right)}{r^{2}}\, \right.   \notag \\
&&\qquad \qquad +\left. 2\Lambda -4b^{2}\left( 1-\frac{1}{\sqrt{1+\frac{q^{2}%
}{b^{2}r^{2D-4}}}}\right) \right] \label{IE_bulk}
\end{eqnarray}
Using the equations of motion (\ref{eom2}) and (\ref{diff_f}), the Euclidean
action becomes a total derivative in the radial coordinate, so that
\begin{equation}
I_{bulk}^{E}=\frac{\beta V(\Sigma _{D-2})}{16\pi G}\,\left. \left(
r^{D-2}f^{\prime }+\frac{4r^{D-2}E\,\phi }{\sqrt{1-\frac{E^{2}}{b^{2}}}}
\right) \right\vert _{r_{+}}^{\infty }\,.  \label{Euclid_bulk}
\end{equation}

In even dimensions $D=2n$, the Euclidean boundary term (\ref{B-even})
evaluated on the black hole solution (\ref{BH}, \ref{f}) is
\begin{eqnarray}
\int\limits_{\partial M}d^{2n-1}x\,B_{2n-1}^{E} &=&2n\beta
V(\Sigma
_{2n-2})\,\lim_{r\rightarrow \infty }r^{2n-2}\sqrt{f}\int\limits_{0}^{1}dt
\,\delta _{\lbrack n_{1}\cdots n_{2n-2}]}^{[m_{1}\cdots m_{2n-2}]}  \notag \\
&&\times K_{\tau }^{\tau }\left( \frac{1}{2}\,R_{m_{1}m_{2}}^{n_{1}n_{2}}-
\left( 2n-1\right) t^{2}K_{m_{1}}^{n_{1}}K_{m_{2}}^{n_{2}}\right) \times
\notag \\
&& \left( \frac{1}{2}
\,R_{m_{3}m_{4}}^{n_{3}n_{4}}-t^{2}K_{m_{3}}^{n_{3}}K_{m_{4}}^{n_{4}}\right)
\cdots \left( \frac{1}{2}
\,R_{m_{2n-3}m_{2n-2}}^{n_{2n-3}n_{2n-2}}-t^{2}K_{m_{2n-3}}^{n_{2n-3}}K_{m_{2n-2}}^{n_{2n-2}}\right) \,.
\end{eqnarray}
Replacing Eqs.(\ref{K}, \ref{R(h)}) and using the integral
\begin{equation}
\int\limits_{0}^{1}ds\,\left[ k-\left( 2n-1\right) t^{2}f\right]
\left( k-t^{2}f\right) ^{n-2}=\left( k-f\right) ^{n-1}\,,
\end{equation}
the boundary term becomes
\begin{equation}
c_{2n-1}\int\limits_{\partial
M}d^{2n-1}x\,B_{2n-1}^{E}=-\frac{\beta \,\ell ^{2n-2}V(\Sigma
_{2n-2})}{16\pi G}\,\left. f^{\prime }\left( f-k\right)
^{n-1}\right\vert ^{r=\infty }\,.
\end{equation}
Therefore, the total Euclidean action $I_{2n}^{E}=I_{bulk}^{E}+c_{2n-1}
\int_{\partial M}d^{2n-1}x\,B_{2n-1}^{E}$ is%
\begin{equation}
I_{2n}^{E}=\frac{\beta V(\Sigma _{D-2})}{16\pi G}\left[ \left. \left(
r^{2n-2}f^{\prime }+\frac{4r^{2n-2}E\,\phi }{\sqrt{1-\frac{E^{2}}{b^{2}}}}
\right) \right\vert _{r_{+}}^{\infty }-\ell ^{2n-2}\left. f^{\prime }\left(
f-k\right) ^{n-1}\right\vert ^{r=\infty }\right] \,.
\end{equation}
The contribution at infinity from the bulk action combines with the boundary
one to produce $\beta$ times the Noether mass, as it may be recognized
from Eq.(\ref{Noether_2n}). Also, it is possible to identify the second term
in the above equation as $-\beta Q\Phi $ using\ Eq.(\ref{limQ}). As
Kounterterm series achieves the cancellation of divergences in the
asymptotic charges, the finiteness of the Euclidean action is ensured for
any static black hole.

Therefore, the term at the horizon corresponds to the black hole entropy
\begin{equation}
S=\frac{V(\Sigma _{D-2})\,r_{+}^{D-2}}{4G}=\frac{\text{Area}}{4G}\,,
\end{equation}
which satisfies the Smarr relation%
\begin{equation}
I_{2n}^{E}=\beta \mathcal{E}-\beta Q\Phi -S\,.  \label{Smarr_2n}
\end{equation}

The energy $\mathcal{E}$ and electric charge $Q$ obtained in
Sec.\ref{conserved} as Noether charges can also be rederived thermodynamically in
the grand canonical ensemble using the definitions (\ref{internal}) and
(\ref{TD_charge}). In order to carry out these calculations, we introduce a new
variable
\begin{equation}
\eta =\frac{q}{br_{+}^{D-2}}\,,  \label{eta}
\end{equation}
instead of $q$. The horizon radius is expressed in terms of the variable $%
\eta $ as
\begin{equation}
r_{+}\left( \eta \right) =\frac{\left( D-3\right) \,\Phi }{b\eta
\,_{2}F_{1}\left( \frac{1}{2},\frac{D-3}{2D-4};\frac{3D-7}{2D-4};-\eta
^{2}\right) }\,,  \label{r(eta,phi)}
\end{equation}
and the energy and electric charge as
\begin{eqnarray}
M\left( \eta \right)  &=&\frac{V(\Sigma _{D-2})\,\left( D-2\right)
\,r_{+}^{D-3}}{16\pi G}\,\left[ k+\frac{r_{+}^{2}}{\ell ^{2}}+
\frac{4b^{2}r_{+}^{2}\left( 1-\sqrt{1+\eta ^{2}}\right) }{\left( D-1\right) \left(
D-2\right) }+\frac{4br_{+}\eta \Phi }{\left( D-1\right) }\right] \,,
\label{M(eta)} \\
Q\left( \eta \right)  &=&\frac{V(\Sigma _{D-2})\,b}{4\pi G}\,\eta
\,r_{+}^{D-2}\,,  \label{Q(eta)}
\end{eqnarray}
while the potential is obtained as the largest root of Eq.(\ref{T}),
\begin{eqnarray}
\Phi (\eta ) &=&-\frac{2b\eta \,_{2}F_{1}\left( \frac{1}{2},\frac{D-3}{2D-4};
\frac{3D-7}{2D-4};-\eta ^{2}\right) }{\left( D-3\right) \left[ \frac{\left(
D-1\right) }{\ell ^{2}}+\frac{4b^{2}}{D-2}\left( 1-\sqrt{1+\eta ^{2}}\right)
\right] }\times   \notag \\
&&\qquad \qquad \times \left[ \frac{\pi }{\beta }+\sqrt{
\frac{\pi ^{2}}{\beta ^{2}}-\left( D-3\right) \left( \frac{\left(
D-1\right) k}{4\ell ^{2}}+
\frac{kb^{2}\left( 1-\sqrt{1+\eta ^{2}}\right) }{D-2}\right) }\right] \,.
\label{phi(eta)}
\end{eqnarray}
After a straightforward calculation, using Eqs.(\ref{r(eta,phi)}-
\ref{phi(eta)}), we find
\begin{eqnarray}
\left( \frac{\partial I^{E}}{\partial \beta }\right) _{\Phi } &=&
\frac{\left( \frac{\partial I^{E}}{\partial \eta }\right) _{\Phi }}{\left(
\frac{\partial \beta }{\partial \eta }\right) _{\Phi }}=\mathcal{E}-Q\Phi \,, \\
\left( \frac{\partial I^{E}}{\partial \Phi }\right) _{\beta } &=&
\frac{\left( \frac{\partial I^{E}}{\partial \eta }\right) _{\beta }}{\left(
\frac{\partial \Phi }{\partial \eta }\right) _{\beta }}=-\beta Q\,.
\end{eqnarray}
Thus, from the above relations, it directly follows that internal energy
(\ref{internal}) is
\begin{equation}
U=\mathcal{E}\,,  \label{energy}
\end{equation}
whereas the thermodynamic charge (\ref{TD_charge}) is given by
\begin{equation}
\mathcal{Q}=Q\,.  \label{thermodynamic_charge}
\end{equation}
It is reassuring to prove that the entropy (\ref{entropy}) is
\begin{equation}
S=\frac{\text{Area}}{4G}\,,
\end{equation}
as expected.


\section{Black hole thermodynamics in odd dimensions}

\subsection{Variational principle and regularization problem}

Three-dimensional AdS gravity can be written as a Chern-Simons density for
the $SO(2,2)$ group. In this formulation, the bulk action comes naturally
supplemented by a boundary term that is half of the Gibbons-Hawking term,
which makes the action finite. This \emph{extrinsic} regularization can be,
therefore, thought as built-in in a Chern-Simons AdS form
\cite{Miskovic-Olea}. In higher odd dimensions, the corresponding Chern-Simons
form for the AdS group produces a gravity theory whose bulk action is given
by Lovelock-type series with specific coefficients. A finite action
principle for this theory is obtained adding a boundary term which is a
given polynomial in the extrinsic and intrinsic curvatures
\cite{Mora-Olea-Troncoso-Zanelli-CS,Mora-Olea-Troncoso-Zanelli-TF}. The same form
of the boundary term solves the regularization problem for Einstein-Hilbert
AdS gravity \cite{Olea-K}. In this case, the Kounterterms series written in
a concise form in terms of the parametric integrations
\begin{eqnarray}
B_{2n}
&=&2n\sqrt{-h}\int\limits_{0}^{1}dt\int\limits_{0}^{t}ds\,\delta
_{\lbrack i_{1}\cdots i_{2n}]}^{[j_{1}\cdots
j_{2n}]}\,K_{j_{1}}^{i_{1}}\delta _{j_{2}}^{i_{2}}\left( \frac{1}{2}
\,R_{j_{3}j_{4}}^{i_{3}i_{4}}-t^{2}K_{j_{3}}^{i_{3}}K_{j_{4}}^{i_{4}}
+\frac{s^{2}}{\ell ^{2}}\,\delta _{j_{3}}^{i_{3}}\delta _{j_{4}}^{i_{4}}\right)
\times  \notag \\
&&\cdots \times \left( \frac{1}{2}
\,R_{j_{2n-1}j_{2n}}^{i_{2n-1}i_{2n}}-t^{2}K_{j_{2n-1}}^{i_{2n-1}}K_{j_{2n}}^{i_{2n}}+
\frac{s^{2}}{\ell ^{2}}\,\delta _{j_{2n-1}}^{i_{2n-1}}\delta
_{j_{2n}}^{i_{2n}}\right) \,.  \label{B_2n}
\end{eqnarray}
With the coupling constant choice
\begin{eqnarray}
c_{2n} &=&-\frac{\ell ^{2(n-1)}}{32\pi G\,n^{2}\left( 2n-1\right) !}\left[
\int\limits_{0}^{1}dt\,\int\limits_{0}^{t}ds\,\left( s^{2}-t^{2}\right)
^{n-1}\right] ^{-1}  \notag \\
&=&\frac{1}{16\pi G}\frac{\left( -1\right) ^{n}\ell ^{2n-2}}{2^{2n-2}n\left(
n-1\right) !^{2}}\,,
\end{eqnarray}
the total action varies on-shell as
\begin{eqnarray}
\delta I_{2n+1} &=&-\frac{1}{2^{n-1}}\int\limits_{\partial M}d^{2n}x\,
\sqrt{-h}\,\delta _{\lbrack i_{1}\cdots i_{2n}]}^{[j_{1}\cdots j_{2n}]}\,\left[
\left( h^{-1}\delta h\right) _{k}^{i_{1}}K_{j_{1}}^{k}+2\delta
K_{j_{1}}^{i_{1}}\right] \delta _{j_{2}}^{i_{2}} \times\notag \\
&&\qquad\qquad\times\left[ \frac{1}{16\pi
G\,(2n-1)!}\,\delta _{\lbrack j_{3}j_{4}]}^{[i_{3}i_{4}]}\cdots \delta
_{\lbrack j_{2n-1}j_{2n}]}^{[i_{2n-1}i_{2n}]}+\right.  \notag \\
&& +\left. nc_{2n}\int\limits_{0}^{1}dt\,
\left( \hat{R}_{j_{3}j_{4}}^{i_{3}i_{4}}+\frac{t^{2}}{\ell ^{2}}\,\delta _{\lbrack
j_{3}j_{4}]}^{[i_{3}i_{4}]}\right) \cdots \left(
\hat{R}_{j2n-1j_{2n}}^{i_{2n-1}i_{2n}}+\frac{t^{2}}{\ell ^{2}}\,\delta _{\lbrack
j_{2n-1}j_{2n}]}^{[i_{2n-1}i_{2n}]}\right) \right]  \notag \\
&&+\frac{nc_{2n}}{2^{n-1}}\int\limits_{\partial
M}\int\limits_{0}^{1}dt\,t\,d^{2n}x\,\sqrt{-h}\,\delta _{\lbrack i_{1}\cdots
i_{2n}]}^{[j_{1}\cdots j_{2n}]}\!\left[ \left( h^{-1}\delta h\right)
_{k}^{i_{1}}\left( K_{j_{1}}^{k}\delta _{j_{2}}^{i_{2}}-\delta
_{j_{1}}^{k}K_{j_{2}}^{i_{2}}\right) +2\delta _{j_{2}}^{i_{2}}\delta
K_{j_{2}}^{i_{2}}\right] \times  \notag \\
&&\times \left(
R_{j_{3}j_{4}}^{i_{3}i_{4}}-t^{2}K_{[j_{3}}^{[i_{3}}K_{j_{4}]}^{i_{4}]}+
\frac{t^{2}}{\ell ^{2}}\,\delta _{\lbrack j_{3}j_{4}]}^{[i_{3}i_{4}]}\right)
\cdots \left(
R_{j_{2n-1}j_{2n}}^{i_{2n-1}i_{2n}}-t^{2}K_{[j_{2n-1}}^{[i_{2n-1}}K_{j_{2n}]}^{i_{2n}]}+
\frac{t^{2}}{\ell ^{2}}\,\delta _{\lbrack
j_{2n-1}j_{2n}]}^{[i_{2n-1}i_{2n}]}\right)  \notag \\
&&\qquad \qquad +\frac{1}{4\pi G}\int\limits_{\partial M}d^{2n}x\,\sqrt{-h}
\,\frac{NF^{ri}\delta A_{i}}{\sqrt{1+\frac{F^{2}}{2b^{2}}}}\,.
\label{varI_2n+1}
\end{eqnarray}
As usual, the surface term from the electromagnetic part vanishes fixing the
gauge potential at the boundary, Eq. (\ref{bc A}). The first two lines of
the above relation can be factorized as
\begin{equation}
-\frac{nc_{2n}}{2^{n-1}}\int\limits_{\partial M}d^{2n}x\,\sqrt{-h}\,\delta
_{\lbrack i_{1}\cdots i_{2n}]}^{[j_{1}\cdots j_{2n}]}\,\left[ \left(
h^{-1}\delta h\right) _{k}^{i_{1}}K_{j_{1}}^{k}+2\delta K_{j_{1}}^{i_{1}}
\right] \delta _{j_{2}}^{i_{2}}\left( \hat{R}_{j_{3}j_{4}}^{i_{3}i_{4}}+
\frac{1}{\ell ^{2}}\,\delta _{\lbrack j_{3}j_{4}]}^{[i_{3}i_{4}]}\right)
P_{j_{5}\cdots j_{2n}}^{i_{5}\cdots i_{2n}}(\hat{R})\,,
\end{equation}
where $P_{j_{5}\cdots j_{2n}}^{i_{5}\cdots i_{2n}}(\hat{R})$ is a
Lovelock-type polynomial of degree $(n-2)$ in the Riemann tensor
$\hat{R}_{kl}^{ij}$ (for explicit expression, see \cite{Olea-K}). Thus, this term is
identically vanishing for the spacetimes with locally AdS asymptotics, i.e.,
satisfying the condition (\ref{bc R}). On the other hand, for any
asymptotically AdS spacetime, the Fefferman-Graham theorem
\cite{Fefferman-Graham} ensures that there exists a regular expansion for the
asymptotic extrinsic curvature, $K_{j}^{i}=-\frac{1}{\ell }\,\delta
_{j}^{i}+\cdots $, where the additional terms are negative powers of $r$ (in
Schwarzschild-like coordinates). This justifies the choice of the asymptotic
condition
\begin{equation}
K_{j}^{i}=-\frac{1}{\ell }\,\delta _{j}^{i}\,,\qquad \text{at }\partial M\,,
\label{K=L}
\end{equation}
which, on top of
\begin{equation}
\delta K_{j}^{i}=0\,,\qquad \text{at }\partial M\,,
\end{equation}
guarantees a well-posed action principle for odd-dimensional AdS gravity
\cite{Olea-K}.

Notice that the asymptotic conditions in both the extrinsic curvature and
the spacetime Riemann tensor are not longer valid for EBI-dilaton theory,
where the solutions are neither asymptotically flat nor (A)dS (see, e.g.,
\cite{Sheykhi}).


\subsection{Conserved quantities}

In odd dimensions, we use Eq.(\ref{varI_2n+1}) and write the surface term
$\Theta $ as
\begin{eqnarray}
\Theta (\xi ) &=&-\frac{1}{2^{n-1}}\,\sqrt{-h}\,\delta _{\lbrack i_{1}\cdots
i_{2n}]}^{[j_{1}\cdots j_{2n}]}\,\left[ \left( h^{-1}\pounds _{\xi }h\right)
_{k}^{i_{1}}K_{j_{1}}^{k}+2\pounds _{\xi }K_{j_{1}}^{i_{1}}\right] \delta
_{j_{2}}^{i_{2}}\left[ \frac{1}{16\pi G\,(2n-1)!}\,\delta _{\lbrack
j_{3}j_{4}]}^{[i_{3}i_{4}]}\cdots \delta _{\lbrack
j_{2n-1}j_{2n}]}^{[i_{2n-1}i_{2n}]}\right.  \notag \\
&&\qquad \qquad +\left. nc_{2n}\int\limits_{0}^{1}dt\,\left(
\hat{R}_{j_{3}j_{4}}^{i_{3}i_{4}}+\frac{t^{2}}{\ell ^{2}}\,\delta _{\lbrack
j_{3}j_{4}]}^{[i_{3}i_{4}]}\right) \cdots \left(
\hat{R}_{j2n-1j_{2n}}^{i_{2n-1}i_{2n}}+\frac{t^{2}}{\ell ^{2}}\,\delta _{\lbrack
j_{2n-1}j_{2n}]}^{[i_{2n-1}i_{2n}]}\right) \right]  \notag \\
&&+\,\frac{nc_{2n}}{2^{n-1}}\int\limits_{\partial
M}\int\limits_{0}^{1}dt\,t\,d^{2n}x\,\sqrt{-h}\,\delta _{\lbrack i_{1}\cdots
i_{2n}]}^{[j_{1}\cdots j_{2n}]}\!\left[ \left( h^{-1}\pounds _{\xi }h\right)
_{k}^{i_{1}}\left( K_{j_{1}}^{k}\delta _{j_{2}}^{i_{2}}-\delta
_{j_{1}}^{k}K_{j_{2}}^{i_{2}}\right) +2\pounds _{\xi
}K_{j_{1}}^{i_{1}}\delta _{j_{2}}^{i_{2}}\right] \times  \notag \\
&&\times \left(
R_{j_{3}j_{4}}^{i_{3}i_{4}}-t^{2}K_{[j_{3}}^{[i_{3}}K_{j_{4]}}^{i_{4}]}+
\frac{t^{2}}{\ell ^{2}}\,\delta _{\lbrack j_{3}j_{4}]}^{[i_{3}i_{4}]}\right)
\cdots \left(
R_{j_{2n-1}j_{2n}}^{i_{2n-1}i_{2n}}-t^{2}K_{[j_{2n-1}}^{[i_{2n-1}}K_{j_{2n}]}^{i_{2n}]}+
\frac{t^{2}}{\ell ^{2}}\,\delta _{\lbrack
j_{2n-1}j_{2n}]}^{[i_{2n-1}i_{2n}]}\right)  \notag \\
&&\qquad \qquad +\frac{1}{16\pi G}\,\sqrt{-h}\,\frac{4NF^{ri}\pounds _{\xi
}A_{i}}{\sqrt{1+\frac{F^{2}}{2b^{2}}}}\,.
\end{eqnarray}
Proceeding similarly as in the even-dimensional case, it can be shown that
the integrand in the Noether charge is split in two pieces, given by
\begin{eqnarray}
q_{i}^{j} &=&-\frac{1}{2^{n-2}}\,\delta _{\lbrack ki_{1}\cdots
i_{2n-1}]}^{[jj_{1}\cdots j_{2n-1}]}\,K_{i}^{k}\delta _{j_{1}}^{i_{1}}\left[
\frac{1}{16\pi G\,(2n-1)!}\,\delta _{\lbrack
j_{2}j_{3}]}^{[i_{2}i_{3}]}\cdots \delta _{\lbrack
j_{2n-2}j_{2n-1}]}^{[i_{2n-2}i_{2n-1}]}\right.  \notag \\
&&\qquad \qquad +\left. nc_{2n}\int\limits_{0}^{1}dt\,\left(
\hat{R}_{j_{2}j_{3}}^{i_{2}i_{3}}+\frac{t^{2}}{\ell ^{2}}\,\delta _{\lbrack
j_{2}j_{3}]}^{[i_{2}i_{3}]}\right) \cdots \left(
\hat{R}_{j_{2n-2}j_{2n-1}}^{i_{2n-2}i_{2n-1}}+\frac{t^{2}}{\ell ^{2}}\,\delta
_{\lbrack j_{2n-2}j_{2n-1}]}^{[i_{2n-2}i_{2n-1}]}\right) \right]  \notag \\
&&\qquad \qquad +\frac{1}{16\pi G}\,\frac{4NF^{rj}\,\left( \xi
^{k}A_{k}\right) }{\sqrt{1+\frac{F^{2}}{2b^{2}}}}\,,
\end{eqnarray}
and
\begin{eqnarray}
q_{(0)i}^{j} &=&\frac{nc_{2n}}{2^{n-2}}\,\delta _{\lbrack ki_{1}\cdots
i_{2n-1}]}^{[jj_{1}\cdots j_{2n-1}]}\,\,\int\limits_{0}^{1}dt\,t\,\left(
K_{i}^{k}\,\delta _{j_{1}}^{i_{1}}+K_{j_{1}}^{k}\delta _{i}^{i_{1}}\right)
\left(
R_{j_{2}j_{3}}^{i_{2}i_{3}}-t^{2}K_{[j_{2}}^{[i_{2}}K_{j_{3}]}^{i_{3}]}+
\frac{t^{2}}{\ell ^{2}}\,\delta _{\lbrack j_{2}j_{3}]}^{[i_{2}i_{3}]}\right)
\times \cdots  \notag \\
&&\qquad \cdots \times \left(
R_{j_{2n-2}j_{2n-1}}^{i_{2n-2}i_{2n-1}}-t^{2}K_{[j_{2n-2}}^{[i_{2n-2}}K_{j_{2n-1}]}^{i_{2n-1}]}+
\frac{t^{2}}{\ell ^{2}}\,\delta _{\lbrack
j_{2n-2}j_{2n-1}]}^{[i_{2n-2}i_{2n-1}]}\right) \,.
\end{eqnarray}%
Therefore, the Noether charge is in odd dimensions is
\begin{equation}
Q[\xi ]=q\left[ \xi \right] +q_{(0)}\left[ \xi \right] \,,  \label{totalQ}
\end{equation}
where
\begin{eqnarray}
q\left[ \xi \right] &=&\int\limits_{\Sigma _{\infty }}d^{2n-1}y\,
\sqrt{\sigma }\,u_{j}\,\xi ^{i}\,q_{i}^{j}\,,  \label{mass} \\
q_{(0)}\left[ \xi \right] &=&\int\limits_{\Sigma _{\infty }}d^{2n-1}y\,
\sqrt{\sigma }\,u_{j}\,\xi ^{i}\,q_{(0)i}^{j}\,,  \label{vacuum energy}
\end{eqnarray}
and whose explicit expressions are
\begin{eqnarray}
q\left[ \xi \right] &=&\frac{1}{2^{n-2}}\int\limits_{\Sigma
_{\infty }}d^{2n-1}y\sqrt{\sigma }\,\delta _{\lbrack ii_{1}\cdots
i_{2n-1}]}^{[jj_{1}\cdots j_{2n-1}]}\,u_{j}\,\xi
^{k}K_{k}^{i}\delta _{j_{1}}^{i_{1}}\left[ \frac{1}{16\pi
G\,(2n-1)!}\,\delta _{\lbrack j_{2}j_{3}]}^{[i_{2}i_{3}]}\cdots
\delta _{\lbrack
j_{2n-2}j_{2n-1}]}^{[i_{2n-2}i_{2n-1}]}\right.  \notag \\
&&\qquad +\left. nc_{2n}\int\limits_{0}^{1}dt\,\left(
\hat{R}_{j_{2}j_{3}}^{i_{2}i_{3}}+\frac{t^{2}}{\ell ^{2}}\,\delta _{\lbrack
j_{2}j_{3}]}^{[i_{2}i_{3}]}\right) \cdots \left(
\hat{R}_{j_{2n-2}j_{2n-1}}^{i_{2n-2}i_{2n-1}}+\frac{t^{2}}{\ell ^{2}}\,\delta
_{\lbrack j_{2n-2}j_{2n-1}]}^{[i_{2n-2}i_{2n-1}]}\right) \right] \,,
\label{q_mass}
\end{eqnarray}
and%
\begin{eqnarray}
q_{(0)}\left[ \xi \right]
&=&-\frac{nc_{2n}}{2^{n-2}}\int\limits_{\Sigma _{\infty
}}d^{2n-1}y\sqrt{\sigma }\,\delta _{\lbrack ii_{1}\cdots
i_{2n-1}]}^{[jj_{1}\cdots j_{2n-1}]}\,u_{j}\,\xi ^{k}\left(
K_{k}^{i}\,\delta _{j_{1}}^{i_{1}}+K_{j_{1}}^{i}\delta
_{k}^{i_{1}}\right)
\times  \notag \\
&&\int\limits_{0}^{1}dt\,t\,\left(
R_{j_{2}j_{3}}^{i_{2}i_{3}}-t^{2}K_{[j_{2}}^{[i_{2}}K_{j_{3}]}^{i_{3}]}+
\frac{t^{2}}{\ell ^{2}}\,\delta _{\lbrack j_{2}j_{3}]}^{[i_{2}i_{3}]}\right)
\times \cdots  \notag \\
&&\qquad \cdots \times \left(
R_{j_{2n-2}j_{2n-1}}^{i_{2n-2}i_{2n-1}}-t^{2}K_{[j_{2n-2}}^{[i_{2n-2}}K_{j_{2n-1}]}^{i_{2n-1}]}+
\frac{t^{2}}{\ell ^{2}}\,\delta _{\lbrack
j_{2n-2}j_{2n-1}]}^{[i_{2n-2}i_{2n-1}]}\right) \,.  \label{q_0}
\end{eqnarray}
Once again, the electromagnetic part of the action does not contribute to
the charge (\ref{totalQ}).

For $\xi ^{i}=\left( 1,\vec{0}\right) $, the charge (\ref{q_mass}) for
topological static black holes (\ref{BH}, \ref{f}) becomes
\begin{equation}
q\left[ \partial _{t}\right] =\frac{V(\Sigma _{2n-1})}{16\pi G}
\,\lim_{r\rightarrow \infty }\,r^{2n-1}f^{\prime }\left[ 1+16\pi G\,n\left(
2n-1\right) !\,c_{2n}\int\limits_{0}^{1}dt\,\left( \frac{k-f}{r^{2}}
+\frac{t^{2}}{\ell ^{2}}\right) ^{n-1}\right] \,.  \label{M2n+1}
\end{equation}
For large $r$, the above integral can be evaluated through the expansion
\begin{equation*}
\int\limits_{0}^{1}dt\,\left( \frac{k-f}{r^{2}}+\frac{t^{2}}{\ell ^{2}}
\right) ^{n-1}=\frac{\left( -1\right) ^{n-1}2^{n-1}\left( n-1\right) !}{\ell
^{2n-2}\left( 2n-1\right) !!}\left[ 1-\frac{\left( 2n-1\right) \ell ^{2}\mu
}{2r^{2n}}\right] +\mathcal{O}\left( r^{-4n+2}\right) \,,
\end{equation*}
so the black hole mass is
\begin{equation}
q\left[ \partial _{t}\right] =M\,,
\end{equation}
given by the expression (\ref{M}). On the other hand, the charge (\ref{q_0})
is
\begin{equation}
q_{(0)}\left[ \partial _{t}\right] =2n\left( 2n-1\right) !c_{2n}V(\Sigma
_{2n-1})\,\lim_{r\rightarrow \infty }\,\int\limits_{0}^{1}dt\,t\,\left( f-
\frac{rf^{\prime }}{2}\right) \left[ k+\left( \frac{r^{2}}{\ell ^{2}}
-f\right) \,t^{2}\right] ^{n-1}\,,  \label{Vac2n+1}
\end{equation}
that, in the limit $r\rightarrow \infty $, represents the vacuum energy
\begin{equation}
E_{vac}=q_{(0)}\left[ \partial _{t}\right] =(-k)^{n}
\frac{V(\Sigma _{2n-1})}{8\pi G}\,\ell ^{2n-2}\,\frac{(2n-1)!!^{2}}{(2n)!}\,.
\end{equation}
This vacuum energy matches the one of Einstein-Hilbert AdS gravity found in
Ref. \cite{Emparan-Johnson-Myers}. In the next section, we show that the
Smarr relation derived from black hole thermodynamics in EBI gravity is
valid only if the internal energy corresponds to the total Noetherian energy
\begin{equation}
\mathcal{E}=Q\left[ \partial _{t}\right] =M+E_{vac}\,,  \label{energy_odd}
\end{equation}
which shifts the\emph{\ }mass coming from background-dependent methods by a
constant.


\subsection{Regularized Euclidean action and Smarr relation}

In odd dimensions $D=2n+1$, the Euclidean boundary term is obtained plugging
in the static metric ansatz (\ref{BH}) into Eq.(\ref{B_2n}), that is
\begin{eqnarray}
\int\limits_{\partial M}d^{2n}x\,B_{2n}^{E} &=&2n\beta V\left(
\Sigma
_{2n-1}\right) \,\lim_{r\rightarrow \infty }\,r^{2n-1}
\sqrt{f}\int\limits_{0}^{1}dt\int\limits_{0}^{t}ds\,\delta _{\lbrack
n_{1}\cdots n_{2n-1}]}^{[m_{1}\cdots m_{2n-1}]}\, \notag \\
&& \times\left[ \left(
K_{\tau }^{\tau }\delta
_{n_{1}}^{m_{1}}+K_{n_{1}}^{m_{1}}\right)
\left( \frac{1}{2}\,R_{n_{2}n_{3}}^{m_{2}m_{3}}
-t^{2}K_{n_{2}}^{m_{2}}K_{n_{3}}^{m_{3}}+
\frac{s^{2}}{\ell ^{2}}\,
\delta _{n_{2}}^{m_{2}}\delta _{n_{3}}^{m_{3}}\right) \right.\notag \\
&&\qquad
+2\left.\left( n-1\right) K_{n_{1}}^{m_{1}}\delta _{n_{2}}^{m_{2}}\left(
-t^{2}K_{\tau }^{\tau }K_{n_{3}}^{m_{3}}+\frac{s^{2}}{\ell ^{2}}\,\delta
_{n_{3}}^{m_{3}}\right) \right]  \notag \\
&&\times\left( \frac{1}{2}
\,R_{n_{4}n_{5}}^{m_{4}m_{5}}-t^{2}K_{n_{4}}^{m_{4}}K_{n_{5}}^{m_{5}}
+\frac{s^{2}}{\ell ^{2}}\,\delta _{n_{4}}^{m_{4}}\delta _{n_{5}}^{m_{5}}\right)
\times  \notag \\
&&\qquad \cdots \times \left( \frac{1}{2}
\,R_{n_{2n-2}n_{2n-1}}^{m_{2n-2}m_{2n-1}}-t^{2}K_{n_{2n-2}}^{m_{2n-2}}K_{n_{2n-1}}^{m_{2n-1}}+
\frac{s^{2}}{\ell ^{2}}\,\delta _{n_{2n-2}}^{m_{2n-2}}\delta
_{n_{2n-1}}^{m_{2n-1}}\right) \,,
\end{eqnarray}
or explicitly, after using the expressions (\ref{K}) and (\ref{R(h)}),
\begin{eqnarray}
\int\limits_{\partial M}d^{2n}x\,B_{2n}^{E} &=&-2n\left(
2n-1\right) !\,\beta V(\Sigma _{2n-1})\lim_{r\rightarrow \infty
}\,\int\limits_{0}^{1}dt\int\limits_{0}^{t}ds\,\left( k-t^{2}f+s^{2}\,
\frac{r^{2}}{\ell ^{2}}\right) ^{n-2}\times  \notag \\
&&  \left[ \frac{rf^{\prime }}{2}\left( k-\left( 2n-1\right)
t^{2}f+s^{2}\,\frac{r^{2}}{\ell ^{2}}\right) +f\left( k-t^{2}f+\left(
2n-1\right) s^{2}\,\frac{r^{2}}{\ell ^{2}}\right) \right] \,.
\end{eqnarray}
Applying the integrals
\begin{eqnarray}
&&\int\limits_{0}^{1}dt\int\limits_{0}^{t}ds\,\left( k-t^{2}f+s^{2}\,
\frac{r^{2}}{\ell ^{2}}\right) ^{n-2}\left( k-\left( 2n-1\right) t^{2}f+s^{2}\,
\frac{r^{2}}{\ell ^{2}}\right)  \notag \\
&=&\int\limits_{0}^{1}dt\left( k-f+t^{2}\,\frac{r^{2}}{\ell ^{2}}\right)
^{n-1}-\int\limits_{0}^{1}dt\,t\,\left( k-t^{2}f+t^{2}\,\frac{r^{2}}{\ell
^{2}}\right) ^{n-1}
\end{eqnarray}
and%
\begin{equation}
\int\limits_{0}^{1}dt\int\limits_{0}^{t}ds\,\left( k-t^{2}f+s^{2}\,
\frac{r^{2}}{\ell ^{2}}\right) ^{n-2}\left( k-t^{2}f+\left( 2n-1\right) s^{2}\,
\frac{r^{2}}{\ell ^{2}}\right) =\int\limits_{0}^{1}dt\,t\,\left(
k-t^{2}f+t^{2}\,\frac{r^{2}}{\ell ^{2}}\right) ^{n-1}\,,
\end{equation}
the surface term becomes
\begin{eqnarray}
\int\limits_{\partial M}d^{2n}x\,B_{2n}^{E} &=&n\left( 2n-1\right)
!\,\beta V(\Sigma _{2n-1})\lim_{r\rightarrow \infty }\left[
r^{2n-1}f^{\prime
}\int\limits_{0}^{1}dt\left( \frac{k-f}{r^{2}}+\frac{t^{2}}{\ell ^{2}}
\right) ^{n-1}\right. +  \notag \\
&&\qquad \qquad +\left. 2\left( f-\frac{rf^{\prime }}{2}\right)
\int\limits_{0}^{1}dt\,t\,\left( k-t^{2}f+t^{2}\,\frac{r^{2}}{\ell ^{2}}
\right) ^{n-1}\right] \,.
\end{eqnarray}%
When the above boundary term is added to the bulk Euclidean action
(\ref{Euclid_bulk}) with a suitable coupling constant,
$I_{2n+1}^{E}=I_{bulk}^{E}+c_{2n}\int_{\partial M}d^{2n}x\,B_{2n}^{E}\,$, the
total contribution coming from radial infinity can be identified with $\beta
\left( M+E_{vac}\right) $, as seen from the two parts of Noether charge
(\ref{M2n+1}) and (\ref{Vac2n+1}). The consistency of the black hole
thermodynamics is therefore verified through the Smarr relation
\begin{equation}
I_{2n+1}^{E}=\beta \left( M+E_{vac}\right) -\beta Q\Phi
-\frac{\text{Area}}{4G}\,,  \label{IE_2n+1}
\end{equation}
for a total energy which includes the zero-point energy $E_{vac}$.

Proceeding in a similar way as in even dimensions, using the thermodynamic
relations (\ref{eta}-\ref{phi(eta)}), it follows%
\begin{eqnarray}
\left( \frac{\partial I^{E}}{\partial \beta }\right) _{\Phi }
&=&M+E_{vac}-Q\Phi \,, \\
\left( \frac{\partial I^{E}}{\partial \Phi }\right) _{\beta } &=&-\beta Q\,,
\end{eqnarray}
from where the internal energy (\ref{internal}) can be computed as
\begin{equation}
U=M+E_{vac}\,,
\end{equation}
and the thermodynamic charge (\ref{TD_charge}) is simply
\begin{equation}
\mathcal{Q}=Q\,.
\end{equation}
This computation verifies for the odd-dimensional case the consistency
between the Noether charges and the corresponding extensive variables in
black hole thermodynamics.

In the extremal case, with solution parameters related by
eq.(\ref{mass ext}), the  black hole temperature vanishes and thus,
the Euclidean action becomes infinite. However, the Gibbs free
energy defined in eq.(\ref{Gibbs}) remains finite even when we take
the extremal black hole limit
\begin{equation}
G=E_{vac}+\frac{V(\Sigma _{D-2})\,}{16\pi G}\left(
2k\,r_{ext}^{D-3}-\mu _{ext}\right)\,,
\end{equation}
where $E_{vac}$ only appears in odd dimensions.  Therefore, the
analysis carried out above still holds, as the thermodynamics is
described by continuous functions which are finite in the extremal
case. In this respect, the EBI AdS system presents
 an analogous behavior as the one of RN-AdS black holes studied in
Ref.\cite{Chamblin-Emparan-Johnson-Myers}. In particular, the black
hole entropy $S$ is still a quarter of the horizon's area.


\section{Conclusions}

Black hole entropy is expected to be due to underlying microscopic degrees
of freedom at the horizon. At a macroscopic level, $S$\ comes from local
properties of the horizon and can be simply computed, e.g., using Wald's
formalism for a given gravity theory \cite{Wald}. However, the interplay
between thermodynamic quantities and the conserved charges at infinity can
be better understood directly from the evaluation of the Euclidean action.

In this spirit, we have checked the validity of the Smarr relation
(\ref{Smarr}) for EBI AdS black holes in all dimensions using
Kounterterms regularization of the action and the conserved charges.
This is particularly important in the odd-dimensional case, where
the method give rise to a non-vanishing vacuum (Casimir) energy
$E_{vac}$ for AdS spacetime, what cannot be observed using any
background-dependent definition of conserved quantities, and whose
formula cannot be worked out in an arbitrary odd dimension in the
standard counterterms approach.

Kounterterms prescription is interesting because it has been proved
to be universal: the explicit form of the boundary terms which
regularize Einstein-Hilbert AdS gravity remains the same when one
couples the Gauss-Bonnet term in higher dimensions than four
\cite{Kofinas-Olea-EGB}. In this way, the correct black hole
thermodynamics can be recovered from a finite Euclidean action.
Remarkably, this is still true for any higher-curvature gravity
theory of the Lovelock family, whenever an AdS branch can be
defined.

There is some evidence that indicates that the full Dirichlet counterterm
series should be generated from Kounterterms by a suitable expansion of the
fields. This has been carried out explicitly for certain Lovelock theories
where the symmetry enhancement permits the integration of intrinsic
counterterms from the variation of the Dirichlet action
\cite{Miskovic-Olea-K}.

This argument strongly supports the idea that a similar proof might be given
for Einstein-Hilbert and Einstein-Gauss-Bonnet AdS gravity.


\section*{Acknowledgments}

The authors thank P. Mora and T. Utino for useful comments. R.O. thanks PUCV
and the organizers of RTN Winter School at CERN, for hospitality during the
completion of this work. O.M. is supported in part by FONDECYT grant
N$^0$ 11070146 and the PUCV through the project N${^0}$
123.797/2007. The work of R.O. is supported by INFN.


\appendix

\section{Conventions \label{Conventions}}

The totally-antisymmetric Kronecker delta of rank $p$ is defined as the
determinant
\begin{equation}
\delta _{\left[ \mu _{1}\cdots \mu _{p}\right] }^{\left[ \nu _{1}\cdots \nu
_{p}\right] }:=\left\vert
\begin{array}{cccc}
\delta _{\mu _{1}}^{\nu _{1}} & \delta _{\mu _{1}}^{\nu _{2}} & \cdots &
\delta _{\mu _{1}}^{\nu _{p}} \\
\delta _{\mu _{2}}^{\nu _{1}} & \delta _{\mu _{2}}^{\nu _{2}} &  & \delta
_{\mu _{2}}^{\nu _{p}} \\
\vdots &  & \ddots &  \\
\delta _{\mu _{p}}^{\nu _{1}} & \delta _{\mu _{p}}^{\nu _{2}} & \cdots &
\delta _{\mu _{p}}^{\nu _{p}}
\end{array}
\right\vert \,.
\end{equation}
A contraction of $k$ indices in the above Kronecker delta produces a delta
of order $p-k$,
\begin{equation}
\delta _{\left[ \mu _{1}\cdots \mu _{k}\cdots \mu _{p}\right] }^{\left[ \nu
_{1}\cdots \nu _{k}\cdots \nu _{p}\right] }\,\delta _{\nu _{1}}^{\mu
_{1}}\cdots \delta _{\nu _{k}}^{\mu _{k}}=\frac{\left( N-p+k\right) !}{
\left( N-p\right) !}\;\delta _{\left[ \mu _{k+1}\cdots \mu _{p}\right] }^{
\left[ \nu _{k+1}\cdots \nu _{p}\right] }\,,\qquad (1\leq k\leq p\leq N)\,,
\end{equation}
where $N$ is the range of indices.

The Riemann tensor in our notation has the form
\begin{equation}
\hat{R}_{\;\;\beta \mu \nu }^{\alpha }=\partial _{\mu }\hat{\Gamma}_{\beta
\nu }^{\alpha }+\hat{\Gamma}_{\sigma \mu }^{\alpha }\,\hat{\Gamma}_{\beta
\nu }^{\sigma }-\left( \mu \leftrightarrow \nu \right) \,,
\end{equation}
in terms of the Christoffel symbol $\hat{\Gamma}_{\alpha \beta }^{\mu }$.
From the above relation it follows that its variation is
\begin{equation}
\delta \hat{R}_{\,\,\beta \mu \nu }^{\alpha }=\nabla _{\mu }\,\delta
\hat{\Gamma}_{\beta \nu }^{\alpha }-\nabla _{\nu }\,\delta \hat{\Gamma}_{\beta
\mu }^{\alpha }\,.
\end{equation}
The scalar curvature can be conveniently written as
\begin{equation}
\hat{R}=\frac{1}{2}\,\delta _{\lbrack \alpha \beta ]}^{[\mu \nu ]}\,g^{\beta
\gamma }\,\hat{R}_{\,\,\gamma \mu \nu }^{\alpha }\,.
\end{equation}

\section{Stokes' theorem \label{StokesTh}}

The Stokes' theorem states
\begin{equation}
\int\limits_{M}d\Omega =\int\limits_{\partial M}\Omega \,,
\end{equation}
where $\Omega $ is a $d$-form defined on a $(d+1)$-dimensional manifold $M$
parameterized by the coordinates $x^{\mu }$,
\begin{equation}
\Omega =\frac{1}{d!}\,\Omega _{\mu _{1}\cdots \mu _{d}}\,dx^{\mu _{1}}\cdots
dx^{\mu _{d}}\,.
\end{equation}
In components, the Stokes' theorem is ($x^{i}$ are the coordinates on
$\partial M$)
\begin{equation}
\int\limits_{M}\partial _{\mu }\Omega _{\mu _{1}\cdots \mu
_{d}}\,dx^{\mu }dx^{\mu _{1}}\cdots dx^{\mu
_{d}}=\int\limits_{\partial M}\Omega _{i_{1}\cdots
i_{d}}\,dx^{i_{1}}\cdots dx^{i_{d}}\,.  \label{Stokes_aux}
\end{equation}
If we choose the metric $g_{\mu \nu }$ with the signature $\left( -,+,\cdots
,+\right) $ in the following way,
\begin{equation}
g_{\mu \nu }dx^{\mu }dx^{\nu }=g_{tt}\,dt^{2}+h_{ij}\text{\thinspace }
dx^{i}dx^{j}\,,  \label{tt}
\end{equation}
so that the boundary $\partial M$ is placed at $x^{0}=t=Const.$, then the
Stokes' theorem (\ref{Stokes_aux}) can be written as
\begin{equation}
\int\limits_{M}d^{d+1}x\,\partial _{\mu }\left( \sqrt{\left\vert
g_{\mu \nu
}\right\vert }V^{\mu }\right) =\int\limits_{\partial M}d^{d}x\,
\sqrt{\left\vert h_{ij}\right\vert }\,n_{\mu }V^{\mu }\,,  \label{Stokes}
\end{equation}
where
\begin{equation}
V^{\mu }=\frac{1}{d!\sqrt{\left\vert g_{\mu \nu }\right\vert }}\,\varepsilon
^{\mu \mu _{1}\cdots \mu _{d}}\,\Omega _{\mu _{1}\cdots \mu _{d}}\,,
\label{V}
\end{equation}
and the outward pointing unit vector normal to $\partial M$ has components
\begin{equation}
n_{\mu }=\left( n_{t},\vec{0}\right) =\left( -\sqrt{-g_{tt}},\vec{0}\right)
\,,\qquad n^{2}=-1\,.
\end{equation}
The $d$-dimensional Levi-Civita symbol is defined as
\begin{equation}
\varepsilon ^{i_{1}\cdots i_{d}}=-\varepsilon ^{ti_{1}\cdots i_{d}}\,.
\end{equation}
If, however, instead of (\ref{tt}) we choose the metric in the form where
$\partial M$ is placed at $x^{1}=r=const$.,
\begin{equation}
g_{\mu \nu }dx^{\mu }dx^{\nu }=g_{rr}\,dr^{2}+h_{ij}\text{\thinspace }
dx^{i}dx^{j}\,,
\end{equation}
then the outward pointing unit vector normal to $\partial M$ in the Stokes'
theorem (\ref{Stokes}) is
\begin{equation}
n_{\mu }=\left( n_{r},\vec{0}\right) =\left( \sqrt{g_{rr}},\vec{0}\right)
\,,\qquad n^{2}=1\,.
\end{equation}
The difference in sign is due to the $d$-dimensional Levi-Civita symbol
defined as
\begin{equation}
\varepsilon ^{i_{1}\cdots i_{d}}=\varepsilon ^{ri_{1}\cdots i_{d}}\,.
\end{equation}

\section{Hypergeometric function \label{hyper}}

The hypergeometric series has the form
\begin{eqnarray}
_{2}F_{1}(a,b;c;z) &=&1+\frac{ab}{c}\,z+\frac{a\left( a+1\right) b\left(
b+1\right) }{2c\left( c+1\right) }\,z^{2}+\cdots  \notag \\
&=&\sum\limits_{p=0}^{\infty }\frac{(a)_{p}(b)_{p}}{(c)_{p}}\,
\frac{z^{p}}{p!}\,,
\end{eqnarray}
where $(a)_{p}$ is a Pochhammer symbol. It converges if $c$ is not a
negative integer (\emph{i}) for all $|z|<1$ and (\emph{ii}) on the unit
circle $|z|=1$ if $\Re e(c-a-b)>0$. An integral representation of the
hypergeometric function is
\begin{equation}
_{2}F_{1}(a,b;c;z)=\frac{\Gamma (c)}{\Gamma (b)
\Gamma (c-b)}\int\limits_{0}^{1}dt\,\frac{t^{b-1}\left( 1-t\right)
^{c-b-1}}{\left( 1-zt\right) ^{a}}\,,\qquad \Re e(c)>\Re e(b)>0\,,
\end{equation}
and its derivative is
\begin{equation}
\frac{d}{dz}\,_{2}F_{1}(a,b;c;z)=\frac{ab}{c}\,_{2}F_{1}(a+1,b+1;c+1;z)\,.
\end{equation}
In particular, the following integral is solved in the text,
\begin{equation}
\int\limits_{0}^{1}dt\,\frac{u^{b-1}}{\sqrt{1+zt}}=\frac{1}{b}
\,_{2}F_{1}\left( \frac{1}{2},b;b+1;-z\right) \,,\qquad b>0\,.
\label{integral}
\end{equation}

\end{document}